\documentclass[aps,prx,twocolumn,superscriptaddress,showpacs]{revtex4-1}
\usepackage{amsmath,amssymb,graphicx}

\usepackage[utf8]{inputenc}
\usepackage[T1]{fontenc}
\usepackage{xcolor}

\IfFileExists{newtxtext.sty}   {\usepackage{newtxtext,newtxmath}}
   {\IfFileExists{stix.sty}
      {\usepackage{stix}}
      {\IfFileExists{mathptmx.sty}
      {\usepackage{mathptmx}}{} } }

\usepackage{textcomp}

\usepackage{bm}

\IfFileExists{siunitx.sty}{\usepackage{booktabs,siunitx}}{}

\pdfoutput=1
\usepackage{color}
\definecolor{LinkColor}{rgb}{0.256,0.439,0.588}
\usepackage{hyperref}
\hypersetup{
   pdfauthor={F. F. Assaad},
   pdftitle={Spin and charge dynamics across de-confined quantum critical points in SU(N) models.},
   colorlinks=true,
   citecolor=LinkColor,
   linkcolor=LinkColor,
   urlcolor=LinkColor
}

\usepackage{pifont}

\usepackage{epstopdf}

\begin{document}

\title{A simple fermionic model of  deconfined phases and phase transitions}

\author{F. F. Assaad }
\affiliation{Institut f\"ur Theoretische Physik und Astrophysik, Universit\"at W\"urzburg, 97074 W\"urzburg, Germany }
\author{ T. Grover}
\affiliation{Department of Physics, University of California at San
Diego, La Jolla, CA 92093, USA}
\affiliation{Kavli Institute for
Theoretical Physics, University of California, Santa Barbara, CA
93106, USA}\date{\today}

\begin{abstract}
Using Quantum Monte Carlo simulations, we study a series of models of fermions coupled to quantum Ising spins on a square lattice with $N$ flavors of fermions per site for $N=1,2$ and $3$. The models have an extensive number of conserved quantities but are not integrable, and have rather rich phase diagrams consisting of several exotic phases and phase transitions that lie beyond Landau-Ginzburg paradigm. In particular, one of the prominent phase for $N>1$ corresponds to $2N$ gapless Dirac fermions coupled to an emergent $\mathbb{Z}_2$ gauge field in its deconfined phase. However, unlike a conventional $\mathbb{Z}_2$ gauge theory, we do not impose the `Gauss's Law' by hand and instead, it emerges due to spontaneous symmetry breaking. Correspondingly, unlike a conventional $\mathbb{Z}_2$ gauge theory in two spatial dimensions, our models have a finite temperature phase transition associated with the melting of the order parameter that dynamically imposes the Gauss's law constraint at zero temperature. By tuning a parameter, the deconfined phase undergoes a transition into a short range entangled phase, which corresponds to N\'eel/Superconductor for $N=2$ and a Valence Bond Solid for $N=3$. Furthermore, for $N=3$, the Valence Bond Solid further undergoes a transition to a N\'eel phase consistent with the deconfined quantum critical phenomenon  studied earlier  in the context of quantum magnets.
\end{abstract}

\pacs{71.10.-w,71.10.Hf,75.40.Cx,75.40.Mg}

\maketitle

\tableofcontents

\section{Introduction}
Ground states of strongly interacting electronic systems can exhibit an extremely rich variety of phases. One way to organize our understanding is to classify the zero temperature phases by their entanglement structure at long distances and low energies \cite{Bisognano75, Bisognano76, Susskind04, Holzhey94, Callan94, Calabrese04, Kitaev06_1, Levin06, Zhang12}. Gapped phases which possess a local order parameter are characterized by short-range entanglement, that is, the reduced density matrix of a large subsystem $A$ can be understood simply by patching together density matrices of smaller subsystems $A_i$ whose union is $A$ \cite{Grover11_2}. This is no longer true for gapless phases such as Fermi liquids \cite{Wolf06,Gioev06,Swingle10}, or gapped topological phases such as a fractional quantum Hall liquid, and such phases are thus said to possess `long-range entanglement' \cite{Kitaev03, Kitaev06_1, Levin06}.

Experience as well as heuristic arguments suggest that Hamiltonians whose ground states posess long range entanglement are relatively difficult to simulate on a classical computer. For example, even a phase as ubiquitous and as well understood as a Fermi liquid is rather hard to simulate numerically because fermions at finite density with  repulsive interactions tend to have an intricate sign structure in their wavefunctions leading to the infamous Monte Carlo fermion sign problem \cite{Hirsch81, Blankenbecler81, Troyer05}. Similarly, fractional quantum Hall phases again possess a non-trivial sign structure to their wavefunctions \cite{Laughlin83}, and therefore so far are amenable only via techniques such as Exact Diagonalization \cite{Laughlin83_2, Yoshioka83} and Density Matrix RG \cite{Feiguin08}, which are restricted to only small two-dimensional systems due to exponential scaling of numerical cost with system size. However, there do exist long-range entangled phases that do not suffer from Monte Carlo sign problem. Two prominent examples are: (1) Interacting Dirac fermions with an even number of flavors \cite{Dagotto89} (2) A gapped $\mathbb{Z}_2$ topological ordered system such as a Toric code   Hamiltonian \cite{Kitaev03} in a magnetic field \cite{Trebst07}. The absence of sign problem for the former is related to the positive fermion determinant, while in the latter case, the Hamiltonian has non-positive off-diagonal elements in a local basis, allowing one to sample the corresponding Boltzmann weight efficiently \cite{Suzuki76}. In this paper, we study a sign problem free model that has a ground state with features of both of the aforementioned long range entangled phases together, namely,  Dirac fermions coupled to a fluctuating $\mathbb{Z}_2$ gauge field.  By tuning parameters in the Hamiltonian, we also study competition with symmetry breaking short-range entangled phases  which leads to novel strongly interacting quantum critical phenomena.

\begin{figure}
\includegraphics[width=0.9\linewidth]{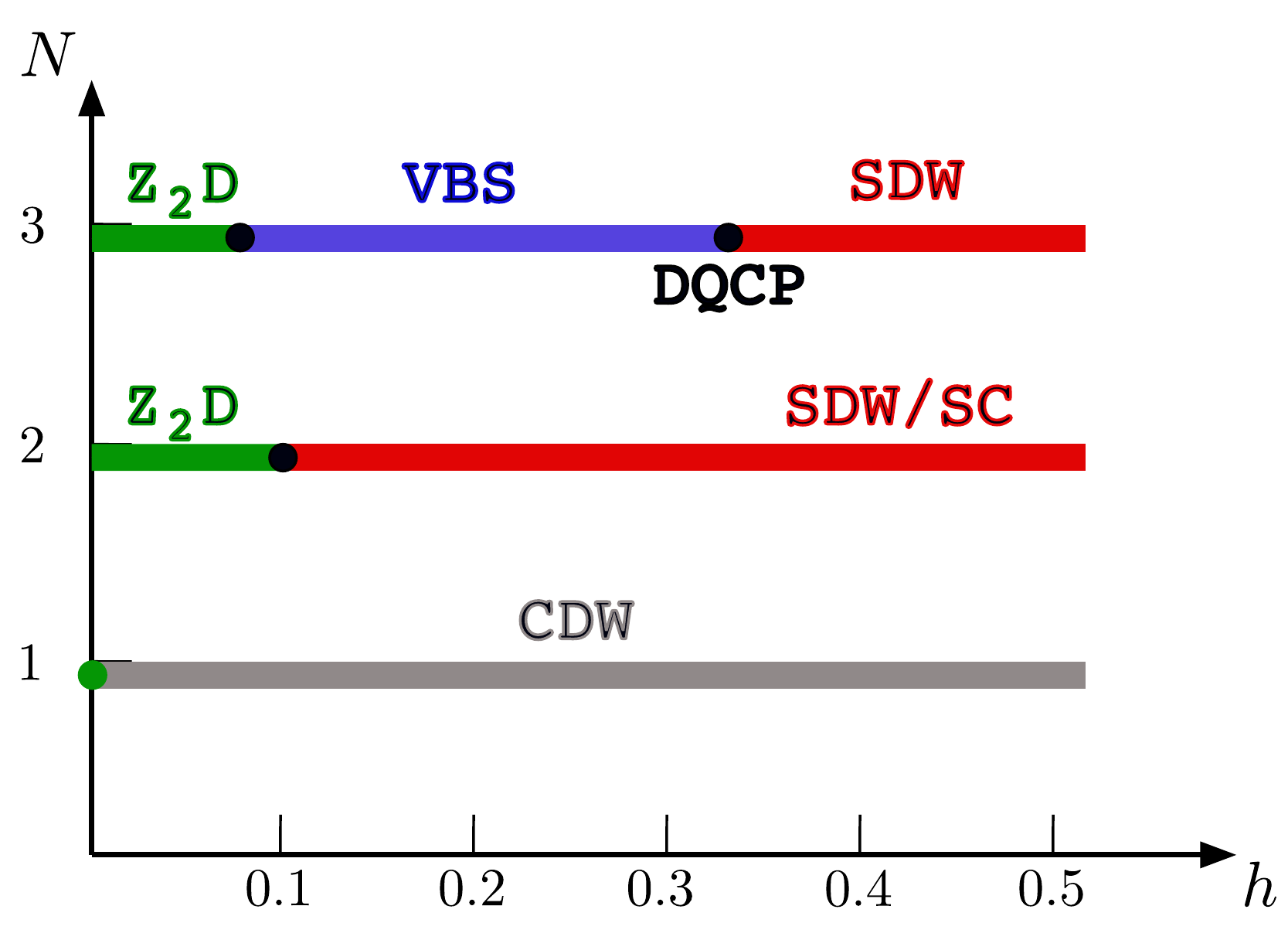}
\caption{(color online)  Schematic zero temperature phase diagram  of the model in Eq.~\ref{Model.Eq}.  We observe a $\mathbb{Z}_2$ Dirac deconfined phase ($Z_2$D), a spin-density wave (SDW) phase ($\equiv$ N\'eel phase) (or a superconductor (SC), depending on the pattern of particle-hole symmetry breaking), a charge density wave (CDW)  phase as well as a valence bond solid (VBS).  For $N=1$, we do not find evidence for a $Z_2$D phase beyond $h=0$ consistent with the arguments in the main text.   The phase transitions from  the $Z_2$D to SDW/SC ($N=2$) and VBS ($N=3$)  are seemingly continuous.   At $N=3$ we observe a deconfined   quantum critical point (DCQP) between the VBS and SDW phases.}    
\label{Phase.fig}
\end{figure}
   
Our Hamiltonian (Eq.\ref{Model.Eq}) is partially motivated by a recent study of nematic instability of a Fermi surface \cite{Berg15} and velocity fluctuations in Dirac metals \cite{Xu16}.   It consists of $N$ species of fermions on the vertices of a square lattice interacting with quantum Ising spins that live on the links of the same square lattice. However, in contrast to the Hamiltonian in Ref.~\cite{Berg15},  our Hamiltonian has a local $\mathbb{Z}_2$  invariance, which leads to $L_x L_y$ conserved operators on a $L_x \times L_y$ lattice. Furthermore, we restrict ourselves to half-filling and thus at low energies, instead of a Fermi surface, we either obtain either no fermions (e.g. a Mott insulator) or Fermi points. These differences completely alter the phase diagram of our model compared to Ref.~\cite{Berg15}.

The phase diagram for various values of $N$ is summarized in Fig.\ref{Phase.fig}. For $N>1$, we find that when the kinetic energy of the aforementioned  Ising spins is small compared to the kinetic energy of the fermions, the ground state resembles deconfined phase of the $\mathbb{Z}_2$ gauge theory coupled to $2N$ Dirac fermions (the $N=1$ case is special in that the deconfined fermions are unstable to infinitesimal interactions). However, as we will discuss in detail, there is an important distinction between this phase and the deconfined phase of a  conventional $\mathbb{Z}_2$ gauge matter theory (Ref.~\cite{Kogut75, Fradkin79, Senthil00}). Namely, the local $\mathbb{Z}_2$ invariance in our model is an \textit{actual} symmetry of the Hamiltonian in contrast to the so-called `gauge symmetry' in gauge-matter theories which just corresponds to \textit{redundancy} in the description of the physical states. This is because we \textit{do not} project the Hilbert space of our Hamiltonian to gauge invariant states, that is, we do not impose the Gauss's law. Instead, in our model, the Gauss's law in the ground state emerges due to spontaneous symmetry breaking as the locally conserved operators order in a certain definite pattern. One physical significance of this is that in contrast to regular $\mathbb{Z}_2$ gauge theory, in our model, equal space, unequal time non-gauge invariant correlation functions  \textit{do not} vanish identically. Further, there is a finite temperature Ising transition corresponding to the restoration of the aforementioned symmetry.

As the kinetic energy of the Ising degrees of freedom is increased, we find that the aforementioned gapless Dirac deconfined phase enters a symmetry broken confined phase. The nature of symmetry breaking depends on the value of $N$. For $N=1$ it corresponds to a charge density wave, while for $N=2$, due to an enlarged symmetry, the symmetry broken phase can correspond to a superconductor or a spin density wave, depending on the sign of infinitesimal symmetry breaking field. The $N=3$ case is even  richer. We find that after exiting the deconfined phase, the model enters a valence bond solid (VBS) phase, and as the tuning parameter controlling the kinetic energy of Ising spins is tuned further, the valence-bond solid quantum melts and yields a  anti-ferromagnet phase. The phase transition between the VBS and N\'eel phase, if second-order, will correspond to a deconfined  quantum critical point \cite{Senthil04_2}, reported earlier in the context of quantum magnets \cite{Sandvik07, Lou09}. Our results are consistent with a second order transition.

\section{Model and symmetries} \label{sec:model}
Our model reads: 
\begin{equation}
	\hat{H}=\sum_{ \left< \pmb{i},\pmb{j} \right> }    \hat{Z}_{\langle \pmb{i},\pmb{j} \rangle}  \left( \sum_{\alpha=1}^{N}   \hat{c}^{\dagger}_{\pmb{i},\alpha} \hat{c}_{\pmb{j},\alpha}  + \text{H.c.} \right)   +  N h \sum_{\left < \pmb{i},\pmb{j} \right> }    \hat{X}_{\langle \pmb{i},\pmb{j} \rangle} \label{Model.Eq}
 \end{equation}
 Here, $\hat{c}^{\dagger}_{\pmb{i},\alpha}$  creates a fermion of  flavor $\alpha$ in a Wannier state centered around lattice site $\pmb{i}$ of a square lattice and the sum runs over bonds of nearest neighbors.    Each bond, $\pmb{b} =  \langle \pmb{i},\pmb{j} \rangle $, accommodates a two level system spanned by the Hilbert space  $ | 0 \rangle_{\pmb{b}} $ and  $ | 1 \rangle_{\pmb{b}} $  and in this basis,   $\hat{Z}_{\langle \pmb{i},\pmb{j} \rangle} $, and  $\hat{X}_{\langle \pmb{i},\pmb{j} \rangle}$, corresponds to  the  Pauli spin matrices $\left( \begin{smallmatrix} 1&0\\ 0&-1 \end{smallmatrix} \right)$ and 
$\left( \begin{smallmatrix} 0&1\\ 1&0 \end{smallmatrix} \right)$ respectively. $h$ tunes the strength of the transverse field. Note that we have set the coefficient of the hopping term to unity so that all energy scales, including the size of $h$ and the temperature $T$, will be measured relative to the hopping term.

 \subsection{Discrete Symmetries} \label{sec:sym_dis}
 
 The model (Eq.\ref{Model.Eq}) leads to a number of symmetries.    The  particle-hole symmetry we consider is
\begin{equation}
	 \hat{P}_{\alpha}^{-1}  \hat{c}^{\dagger}_{\pmb{i}, \beta} \hat{P}_{\alpha} = \delta_{\alpha,\beta} e^{i \pmb{K}\cdot \pmb{i}}   \hat{c}_{\pmb{i}, \beta }   + \left(1 -  \delta_{\alpha,\beta} \right) \hat{c}^{\dagger}_{\pmb{i}, \beta}  \label{eq:ph}
\end{equation}
and it satisfies
\begin{equation}
 \left[ \hat{P}_{\alpha}, \hat{H} \right]  = 0. 
 \end{equation}
As an explicit operator, $\hat{P}_{\alpha}   = \prod_i (c_{\pmb{i},\alpha} e^{i \pmb{K}\cdot \pmb{i}/2}  + \hat{c}^{\dagger}_{\pmb{i},\alpha} e^{-i \pmb{K}\cdot \pmb{i}/2}  )$ where $\bm{K}   =  \left( \pi, \pi \right) $. Thus it is a fermionic (bosonic) operator if the total number of lattice sites is odd (even). As it will become clearer later in the paper, it is helpful to define a $\mathbb{Z}_2$ valued order-parameter $\hat{p}_{\pmb{i}}$ that captures the breaking of the particle-hole symmetry,

\begin{equation}
\hat{p}_{\pmb{i}}    = \prod_{\alpha=1}^{N} \left(1 - 2\hat{n}_{\pmb{i},\alpha} \right)   =  \left(- 1\right)^{\hat{N}_{\pmb{i}}} \label{eq:parity}
\end{equation}
This is nothing but the parity of fermion particle number at site $\pmb{i}$. Under particle-hole transformation, $\hat{p}_{\pmb{i}}  \rightarrow - \hat{p}_{\pmb{i}} $.

A more interesting symmetry corresponds to the local conservation laws:

\begin{equation}
 	 \left[\hat{Q}_{\pmb{i}}, \hat{H} \right]  = 0
 \end{equation} 

where 

 \begin{equation}
 	\hat{Q}_{\pmb{i}}  =   \hat{X}_{\pmb{i},\pmb{i} + \pmb{a}_x} \hat{X}_{\pmb{i},\pmb{i} - \pmb{a}_x} \hat{X}_{\pmb{i},\pmb{i} + \pmb{a}_y} \hat{X}_{\pmb{i},\pmb{i} - \pmb{a}_y} \hat{p}_{\pmb{i}}  \label{eq:Q}
 \end{equation}
with $ \left[\hat{Q}_{\pmb{i}}, \hat{Q}_{\pmb{j}} \right]=  0$ and $\hat{Q}_{\pmb{i}}^2= 1$. Despite an extensive number of conservation laws, our Hamiltonian is \textit{not} integrable. This is because the total number of degrees of freedom, $N_{dof}$, after taking into account $L_x L_y$ number of conserved  $\hat{Q}_{\pmb{i}}$'s on a $L_x \times L_y$ lattice, are still extensive: $N_{dof} =$ Number of Ising spins $+$ Number of fermions $-$ Number of constraints $= 2 L_x L_y + N L_x L_y - L_x L_y = (N+1) L_x  L_y$.

Crucially, the operators $\hat{Q}_{\pmb{i}}$ anti-commute with the generator of  particle-hole symmetry: $  \hat{P}^{-1}_{\alpha} \hat{Q}_{\pmb{i}}  \hat{P}_{\alpha} = -\hat{Q}_{\pmb{i}}  $. Therefore, the low energy theory of our model cannot contain terms that are proportional to $\hat{Q}_{\pmb{i}}$ unless the  particle-hole symmetry is spontaneously broken. The anti-commutation also implies that the  energy eigenstates cannot simultaneously be  particle-hole symmetric and eigenstates of $\hat{Q}_{\pmb{i}}$. In particular, particle-hole symmetric eigenstates satisfy $  \langle \hat{Q}_{\pmb{i}}   \rangle = 0 $. However, it's worth emphasizing that even if an energy eigenstate is not particle-hole symmetric, it does not necessarily break the particle-hole symmetry \textit{spontaneously}. For example, one can in principle construct a state $|\psi\rangle$ which is an eigenstate of  all $\hat{Q}_{\pmb{i}}$'s, (and therefore cannot be eigenstate of  $\hat{P}_{\alpha}$ for any $\alpha$), but nevertheless satisfies $\langle \psi|\hat{p}_{\pmb{i}} |\psi\rangle = 0$, and $\langle \psi|\hat{p}_{\pmb{i}} \hat{p}_{\pmb{j}}|\psi\rangle \rightarrow 0$ as $|{\pmb{i}} - {\pmb{j}}| \rightarrow \infty$. 

Another consequence of anticommutation between $\hat{Q}_{\pmb{i}}$ and $\hat{P}_{\alpha}$ is that the spectra of the Hamiltonian is at least doubly degenerate. In fact, on a lattice with an odd number of total sites, the Hamiltonian possesses a $\mathcal{N} = 2$ SUSY \cite{Hsieh16}. This is  because, as discussed in $\hat{P}_{\alpha}$ is a fermionic operator and together with bosonic operator $\hat{Q}_{\pmb{i}} $ can be used to define a fermionic SUSY generator $\mathcal{Q} = \sqrt{H/2} \hat{P}_{\alpha} (1+ \hat{Q}_{\pmb{i}})$ for any  $\alpha$ and $\bm{i}$, so that $\{\mathcal{Q}, \mathcal{Q}^{\dagger}\} = 2 H$, $\mathcal{Q}^2 = 0$, and $[\mathcal{Q},H] = 0$.
 
 At this point it is important to pause and note that even though our Hamiltonian (Eqn.\ref{Model.Eq}) bears a strong resemblance to a $\mathbb{Z}_2$ gauge theory coupled to matter fields \cite{Kogut75, Fradkin79, Senthil00}, ours is actually not a gauge theory due to an important difference. Unlike a matter-gauge theory, we \textit{do not} impose the Gauss's law constraint at any site. In a conventional $\mathbb{Z}_2$ gauge theory, the field  $\hat{X}_{\pmb{i},\pmb{i} + \hat{n}}$ would correspond to the $\mathbb{Z}_2$ electric field along direction $\hat{n}$ at site $\pmb{i}$, and thus the Gauss's law (`$\bm{\nabla}.\bm{E} = \rho$') reads: $ \hat{X}_{\pmb{i},\pmb{i} + \pmb{a}_x} \hat{X}_{\pmb{i},\pmb{i} - \pmb{a}_x} \hat{X}_{\pmb{i},\pmb{i} + \pmb{a}_y} \hat{X}_{\pmb{i},\pmb{i} - \pmb{a}_y}\, ( \equiv \bm{\nabla}.\bm{E}) = \hat{p}_{\pmb{i}}\, (\equiv \rho)$. Satisfying this operator equation would thus require projecting the Hamiltonian to a specific set of eigenvalues of operators $\hat{Q}_{\pmb{i}}$ (namely, $\hat{Q}_{\pmb{i}}=1$ \footnote{In general, one can also allow for the possibility of a background  static $\mathbb{Z}_2$ charge $q_{\textrm{static},\pmb{i}}$ so that the constraint is $\hat{X}_{\pmb{i},\pmb{i} + \pmb{a}_x} \hat{X}_{\pmb{i},\pmb{i} - \pmb{a}_x} \hat{X}_{\pmb{i},\pmb{i} + \pmb{a}_y} \hat{X}_{\pmb{i},\pmb{i} - \pmb{a}_y} = q_{\textrm{static},\pmb{i}} \,\hat{p}_{\pmb{i}}$}). In contrast, we do not impose any constraint on $\hat{Q}_{\pmb{i}}$'s, which therefore behave like classical degrees of freedom, whose values are allowed to vary from one eigenstate to another, and for a given many-body eigenstate these values will be determined by the intrinsic dynamics of our Hamiltonian.
 
The local  conservation of  $\hat{Q}_{\pmb{i}}$ has important consequences.  Since $\hat{Q}_{\pmb{i}}  \hat{c}^{\dagger}_{\pmb{j},\alpha}  {\hat{Q}_{\pmb{i}}}^{-1} = \left(1 - 2 \delta_{ \pmb{i},  \pmb{j}}\right) \hat{c}^{\dagger}_{\pmb{j},\alpha} $, one can readily show that 
\begin{equation}
\label{Green_loc.eq}
  \langle  \hat{c}^{\dagger}_{ {\pmb i}_1,\alpha }(\tau)     \hat{c}_{{\pmb i}_n, \alpha  }(0)   \rangle    = \delta_{\pmb{i}_1,\pmb{i}_n} \langle  \hat{c}^{\dagger}_{{\pmb i}_1, \alpha }(\tau)     \hat{c}_{{\pmb i}_1,\alpha }(0)   \rangle. 
\end{equation}
or  equivalently 
\begin{equation}
  \langle  \hat{Z}_{ {\pmb b} } (\tau)    \hat{Z}_{  {\pmb b}' }  (0)   \rangle    = \delta_{\pmb{b},\pmb{b}'}    \langle  \hat{Z}_{ {\pmb b} } (\tau)    \hat{Z}_{  {\pmb b} }  (0)   \rangle.
 \end{equation}
Thus, a bare single fermion  $\hat{c}_{\pmb{i},\alpha}$ as well as the Ising operator $ \hat{Z} $,  are localized along the real space axis, but can propagate along the imaginary time direction.  Propagation along the imaginary time implies  transitions between different $\hat{Q}_{\pmb{i}}$ sectors which is not prohibited in our model. As already mentioned, if one were to instead impose  a constraint on the values of $\hat{Q}_{\pmb{i}}$, it would prohibit  propagation along the time direction as well, leading to a local  $\mathbb{Z}_2$ gauge redundancy.  

On the other hand, quantities such as 
\begin{equation}
	 \langle  \hat{c}^{\dagger}_{{\pmb i}_1,\alpha }(\tau)  \prod_{j=1}^{n-1} \hat{Z}_{\langle \pmb{i}_j,\pmb{i}_{j+1} \rangle}\left( \tau_{j} \right)  \hat{c}_{\pmb{i}_{n}, \alpha  }(0)   \rangle	
\end{equation}
where one attaches a string  of $\hat{Z}$-operators along  certain chosen bonds connecting $\pmb{i}_1$  to $\pmb{i}_n$,  as well as closed paths of $\hat{Z}$-operators, or  the expectation value of  $  \hat{X}_{\langle \pmb{i}, \pmb{j}\rangle } $  are  all quantities which do not  vanish  due to symmetry arguments.     Again, since the Hilbert space is not restricted to a single choice of $\hat{Q}_i$   the strings are continuous in space but can be discontinuous in the imaginary time.

\subsection{Continuous Symmetries} \label{sec:sym_cont}
 
 The model is clearly invariant under global U($N$)  rotations in flavor space: 
 \begin{equation}
 	\hat{c}^{\dagger}_{\pmb{i},\alpha} \rightarrow \sum_{\beta=1}^{N} U_{\alpha,\beta} \hat{c}^{\dagger}_{\pmb{i},\beta}
 \end{equation}
 where $U$ corresponds to an U($N$) matrix. We recall that U($N$) = $\frac{ \textrm{SU}(N) \times \textrm{U}(1)}{Z_N}$ where the SU($N$) component corresponds to the rotations among the $N$-flavors and the U(1) corresponds to the global particle number conservation $c_{\bm{i},\alpha} \rightarrow e^{i \theta} c_{\bm{i},\alpha}$.  

In fact, the full set of continuous symmetries is bigger than U($N$) and corresponds to O(2$N$). This symmetry is manifest if one makes a unitary transformation $c_{\pmb j,\alpha} \rightarrow i c_{\pmb j,\alpha}$ on only one of the sublattices of the square lattice, and then writes the complex fermions in terms of their Majorana components: $c_{\pmb j,\alpha} = \left(\gamma_{\pmb j,\alpha,1} + i \gamma_{\pmb j,\alpha,2}\right)/2$. The Hamiltonian then becomes:

 \begin{equation}
	\hat{H}=\sum_{\left< \pmb{i},\pmb{j} \right> }  \hat{Z}_{\langle \pmb{i},\pmb{j} \rangle}  \left( \sum_{\alpha=1}^{N}  \sum_{a=1,2} i \gamma_{\pmb i,\alpha,a} \gamma_{\pmb j,\alpha,a} \right)   +  N h \sum_{\left< \pmb{i},\pmb{j} \right> }    \hat{X}_{\langle \pmb{i},\pmb{j} \rangle} \label{ModelMaj.Eq}
 \end{equation}
 which is manifestly O(2$N$) invariant.
 
 Since we are in two spatial dimensions, the aforementioned global continuous  symmetry can break spontaneously at zero temperature.   In many of the examples we will discuss, the symmetry breaking will  be manifest in long-ranged correlation functions such as: 
 \begin{equation}
 	S_{\text{Spin}} (\pmb{i} -  \pmb{j}, \tau -  \tau')    =   \frac{1}{N} \sum_{\alpha,\beta} \langle  \hat{S}^{\alpha}_{ \; \; \beta}(\pmb{i}, \tau)   \hat{S}^{\beta}_{ \; \; \alpha}(\pmb{j}, \tau') \rangle. 
 \end{equation}  
 Here 
 \begin{equation}
 \hat{S}^{\alpha}_{ \; \; \beta}(\pmb{i})  = \hat{c}^{\dagger}_{\pmb{i},\alpha} \hat{c}^{}_{\pmb{i},\beta}    - \frac{1}{N}  \delta_{\alpha,\beta} \sum_{\gamma = 1}^N \hat{c}^{\dagger}_{\pmb{i},\gamma} \hat{c}^{}_{\pmb{i},\gamma} 
  \end{equation}
  are the generators of the subgroup SU(N) of the aforementioned O(2$N$) symmetry.

Before moving on, we note that the discrete particle-hole symmetry provides exact mapping between seemingly different correlations functions. For example,  let us consider the  transverse  ($\alpha \neq \beta$) spin operator  
\begin{equation}
	 \hat{P}^{-1}_{\beta} \hat{S}^{\alpha}_{ \; \; \beta}(\pmb{i}, \tau)  \hat{P}_{\beta} =  e^{i\pmb{K} \cdot \pmb{i} } \hat{c}^{\dagger}_{\pmb{i}, \alpha} (\tau) \hat{c}^{\dagger}_{\pmb{i}, \beta} (\tau). 
\end{equation}
Hence, the particle-hole condensate at  $\pmb{q} =\pmb{K} = \left( \pi, \pi \right) $  is equivalent to a particle-particle condensate at $\pmb{q} =  \left( 0,0 \right) $. Equivalently, a charge density wave  at  $\pmb{q}   =  \left( \pi, \pi \right) $ transforms to a  diagonal spin-density wave at the same wave vector. 

\section{Effective model at weak and strong coupling}

 \subsection{Strong transverse field limit} \label{sec:strong}
 
 In this limit, we  can understand  the model in terms of a Su-Schrieffer-Heeger coupling \cite{Su80} to an Einstein phonon mode of frequency $h$,  but with a truncated phonon Hilbert space.  This reading of the model is certainly valid in the high frequency limit where phonon modes are  not highly occupied.  In particular,  and in the 
 $ | \pm \rangle = \frac{1}{\sqrt{2}}  \left(  | 1 \rangle + | 0 \rangle \right)  $ 
 basis,    we can represent the Ising spins in  terms of hard core bosons:
 \begin{equation}
 	 \hat{X}_{\pmb{b}}  =  2\hat{b}^{\dagger}_{\pmb{b}} \hat{b}^{}_{\pmb{b}}  - 1,  \; \; \; \; \hat{Z}_{\pmb{b}}  =  \hat{b}^{\dagger}_{\pmb{b}} +  \hat{b}_{\pmb{b}} 
 \end{equation}
At high frequencies,  the occupation of the bosonic level will be small so that   one can safely relax the hard-core boson constraint, $ \left( \hat{b}^{\dagger}_{\pmb{b}}  \right)^2 =0$, and   promote the bosonic operators to soft core ones.   In this approximation, the analogy to phonons is exact. Integrating out the phonons leads to a retarded interaction   resulting in the action:
\begin{equation}
   S  =   - \frac{1}{2 N  h} \sum_{ \langle \pmb{i},\pmb{j} \rangle }  \int_{0}^{\beta}  {\text d}  \tau {\text d}  \tau'    \hat{k}_{\langle \pmb{i},\pmb{j} \rangle }(\tau) 
 D(\tau-\tau')     \hat{k}_{\langle \pmb{i},\pmb{j} \rangle }(\tau')
\end{equation}
with
\begin{equation}
  \hat{k}_{\langle \pmb{i},\pmb{j} \rangle} =  \sum_{\alpha=1}^{N}   \left( \hat{c}^{\dagger}_{\pmb{i},\alpha} \hat{c}_{\pmb{j},\alpha}  + \text{H.c.}  \right),
\end{equation}
the local hopping term,  and 
\begin{equation}
	D(\tau)  = h   \frac{ e^{- 2 \left(  \beta - | \tau|  \right)  h }  + e^{- 2| \tau |   h  } }{1 - e^{- 2 \beta h}},
\end{equation}
the bosonic propagator for $ -\beta < \tau < \beta $.  In the high frequency limit and for constant values of $\beta$   the interaction becomes instantaneous  $ \lim_{h \rightarrow \infty} D(\tau) = \delta( \tau) $   and  the full Hamiltonian reduces to 
\begin{equation}
\label{h_inf.eq}
	\hat{H}_{\infty}  =       - \frac{1}{2 N  h} \sum_{ \left< \pmb{i},\pmb{j} \right> }  \hat{k}^2_{\langle \pmb{i},\pmb{j} \rangle}.
\end{equation}
This interaction was recently considered in the context of SU($N$) fermions on the honeycomb lattice in Ref. \cite{Li15a}.  

The form in Eq.~\ref{h_inf.eq}  is easy to understand.  In the strong coupling limit,   the Ising spins are polarized  along the $x$-direction, and fermion hopping is prohibited since this implies  flipping of the spin against the transverse field.  Virtual processes can nevertheless occur and  generate  a super exchange  energy scale $J \propto  \frac{1}{2 N  h} $ .   

It is instructive to consider the nature of  phases in this limit for a few values of $N$. First, at large $N$, the Hamiltonian $\hat{H}_{\infty}$ in Eq.\ref{h_inf.eq} can be solved exactly (Ref.\cite{Affleck88}) and the ground state corresponds to a valence bond solid state. For $N=1$ case, the Hamiltonian is simply $H = \frac{1}{2 N  h} \sum_{ \langle \pmb{i},\pmb{j} \rangle } \left(n_{ \pmb{i}} - \frac{1}{2}\right)  \left(n_{ \pmb{j}} - \frac{1}{2}\right)$ and thus the ground state corresponds to a charge-density wave. For $N=2$ case, one finds   
\begin{equation}
	 - \frac{1}{4 h} \sum_{ \left< \pmb{i},\pmb{j} \right> }  \hat{k}^2_{\langle \pmb{i},\pmb{j} \rangle}  = 
	  \frac{1}{ h} \sum_{ \left< \pmb{i},\pmb{j} \right> } \left(   \hat{\pmb{S}}_{\pmb{i}}   \hat{\pmb{S}}_{\pmb{j}}    +   \hat{\pmb{\eta}}_{\pmb{i}}   \hat{\pmb{\eta}}_{\pmb{j}}   \right). \label{eq:Hstrongh}
\end{equation}
Here $\hat{\pmb{S}}_{\pmb{i}}  $ is the spin-1/2 operator and  
$\hat{\pmb{\eta}}_{\pmb{i}}  = \hat{P}^{-1}_{\uparrow} \hat{\pmb{S}}_{\pmb{i}} \hat{P}_{\uparrow}  $ is the `Anderson's pseudospin' \cite{Anderson58},\footnote{\label{footnote:bipart} Note that on a general lattice, the second term would instead be $\eta_i^{z} \eta_j^{z} - \frac{1}{2}\left(\eta_i^{+} \eta_j^{-} + \eta_i^{-} \eta_j^{+}\right)$. On a bipartite lattice, such as ours, one can do a unitary transformation to bring it to the form in Eq.\ref{eq:Hstrongh}}.
In the strong coupling limit  charge is localized and the Ising spins are frozen in the $x$-direction  wave functions.  Such functions have well defined quantum numbers  $\hat{Q}_{\pmb{i}}$.  If one choses $\hat{Q}_{\pmb{i}}  = -1$, then the ground state will be a  N\'eel state for the fermions while the Ising spins point along the x-direction. It is interesting to note that at $h = \infty$  the kinetic energy vanishes and that all real space charge configurations  and thereby all $\hat{Q}_{\pmb{i}}  $ sectors are degenerate.

\subsection{Weak transverse field limit } \label{sec:weak}
In the $h \rightarrow 0$   limit,    the imaginary time scale  on which the Ising spins can fluctuate diverges such  that  the Ising spin degrees of freedom become classical.    The model then reduces to 
\begin{equation}
	\hat{H}_0=\sum_{\left< \pmb{i},\pmb{j} \right> }     Z_{\langle \pmb{i},\pmb{j} \rangle}  \left( \sum_{\alpha=1}^{N}   \hat{c}^{\dagger}_{\pmb{i},\alpha} \hat{c}_{\pmb{j},\alpha}  + \text{H.c.} \right)  
 \end{equation}
 and  one has to find an arrangement of Ising spins which will minimize the energy.  Local $\mathbb{Z}_2$ invariance implies  that  the energy of an eigenstate depends only on the $\mathbb{Z}_2$ flux  $\hat{F}_{\pmb{i}}$ through a plaquette:
 \begin{eqnarray} 
 	 \hat{F}_{\pmb{i}} = & &   \hat{Z}_{\langle \pmb{i},\pmb{i} + \pmb{a}_x \rangle} \hat{Z}_{\langle \pmb{i} + \pmb{a}_x,\pmb{i} +  \pmb{a}_x+ \pmb{a}_y \rangle}  \times \nonumber \\
	            & & \hat{Z}_{\langle \pmb{i} + \pmb{a}_x+ \pmb{a}_y  , \pmb{i} + \pmb{a}_y  \rangle} \hat{Z}_{\langle \pmb{i} + \pmb{a}_y , \pmb{i} \rangle}.
\label{Z2_flux.eq}
\end{eqnarray} 
Lieb's work \cite{Lieb94} implies that the minimal energy is realized by $\pi$-flux configurations i.e. $\hat{F}_{\pmb{i}} = -1$. This means that the coupling  to the fermions effectively generates a  $\mathbb{Z}_2$ {\it magnetic field} term 
\begin{equation}
\label{MF.eq}
	|J| \sum_{\pmb{i}}\hat{F}_{\pmb{i}}   
\end{equation}
in the Hamiltonian. \begin{figure}
\includegraphics[width=0.9\linewidth]{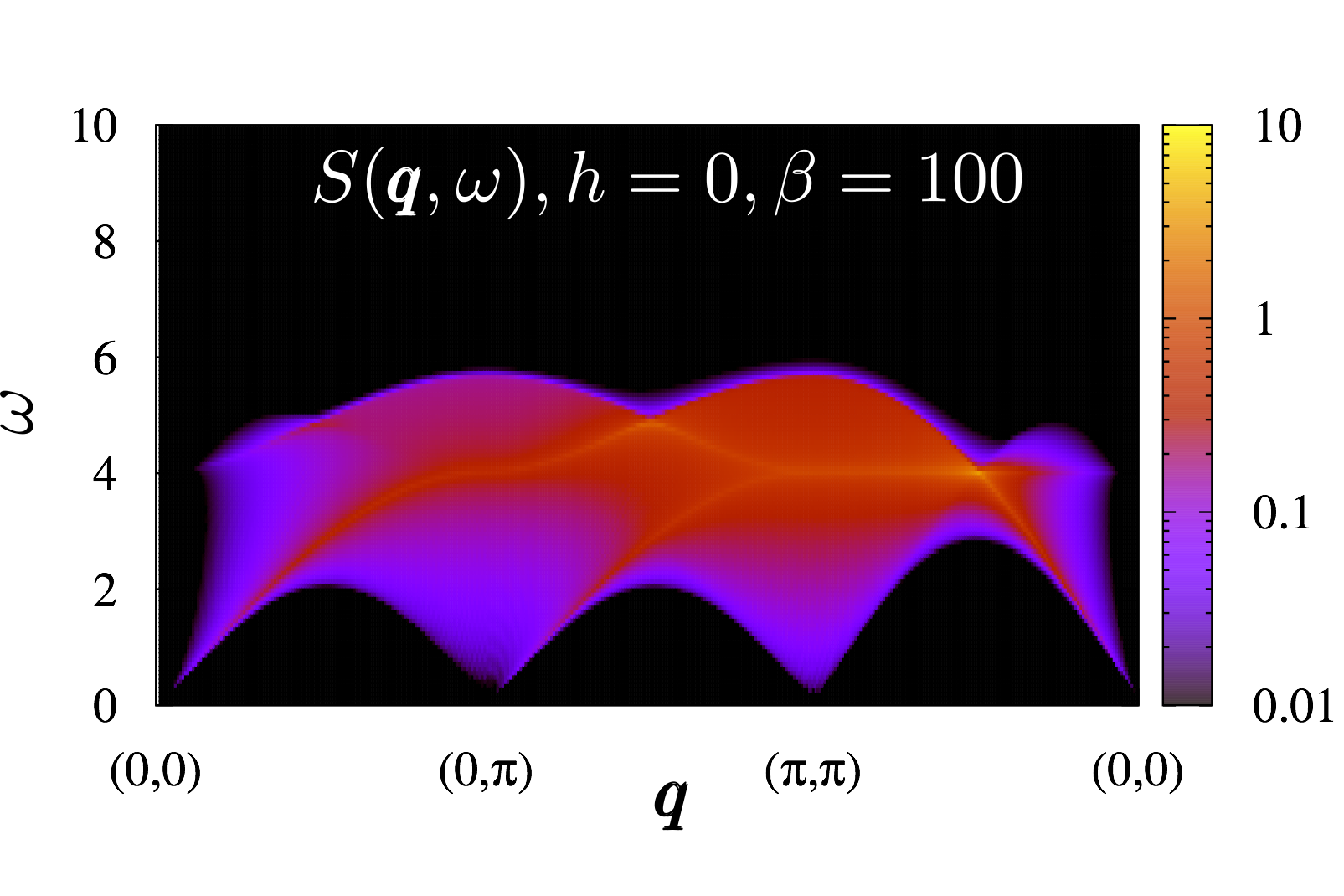}
\caption{(color online)   Dynamical spin structure factor  at $h=0$.   At the considered temperature we can safely neglect the  thermal fluctuations of the Ising fields. Only Ising field configurations with a $\pi$-flux per plaquette contribute. }
\label{Sqom_h0.fig}
\end{figure} \noindent
On the torus with $L_x \times L_y$ sites, there are $2^{L_x L_y}$ configurations of the Ising field that correspond to the $\pi$-flux through every plaquette, and therefore, an equal number of ground states at $h=0$. This huge ground state degeneracy can also be thought of as a result of the conserved quantities $\hat{Q}_{\pmb{i}}$ that relate different ground states. To see the effect of $\pi$-flux on fermions, let us chose a two orbital unit cell  with unit vectors $\pmb{a}_{\pm}  = \pmb{a}_{x} \pm  \pmb{a}_{y} $ such that the  intra-unit cell Ising field is $-1$ while all others are set to $1$.  As mentioned above, the single particle spectral function depends on the specific choice of Ising fields. On the other hand, the two particle  Green functions in the particle-hole  channel  are independent of the background Ising configuration. In Fig.~\ref{Sqom_h0.fig}, we plot the ground  state dynamical    spin-spin correlations function, $S(\pmb{q},\omega) $,  valid for all $N$.  One observes a  {\it particle-hole} continuum with gapless excitations  at wave vectors connecting the Dirac cones $\pmb{q}=(\pi,0)$ and  $\pmb{q}=(\pi,\pi)$.   One key interest of the QMC simulations will be to study the fate of this {\it particle-hole}  continuum at non-zero values of $h$.

As one might expect, and in accord with the third law of thermodynamics, quantum fluctuations will lift the zero temperature finite  entropy density. In general, at non-zero values of $h$ terms which do not violate symmetries of the model will be dynamically generated.  In particular, alongside the magnetic field term of Eq.~\ref{MF.eq}  one can write down,
\begin{equation}
\label{Ising.eq}
	\hat{H}_{Q}  = \sum_{\pmb{i},\pmb{j}}  K_{\pmb{i},\pmb{j}} \hat{Q}_{\pmb{i}} \hat{Q}_{\pmb{j}}.
\end{equation}
which is invariant under the particle-hole symmetry that sends $\hat{Q}_{\pmb{i}}  \rightarrow - \hat{Q}_{\pmb{i}} $. Since this is nothing but a classical Ising Hamiltonian, one can foresee that there will likely be a finite temperature  Ising transition  below which the 
$\hat{Q}_{\pmb{i}} $'s spontaneously order.  As discussed in the following sections, this symmetry breaking will result in an emergent $\mathbb{Z}_2$ gauge structure in the ground state phase diagram and would also determine the nature of both confined and deconfined phases.

\section{Methodology}

\begin{figure}
\includegraphics[width=0.9\linewidth]{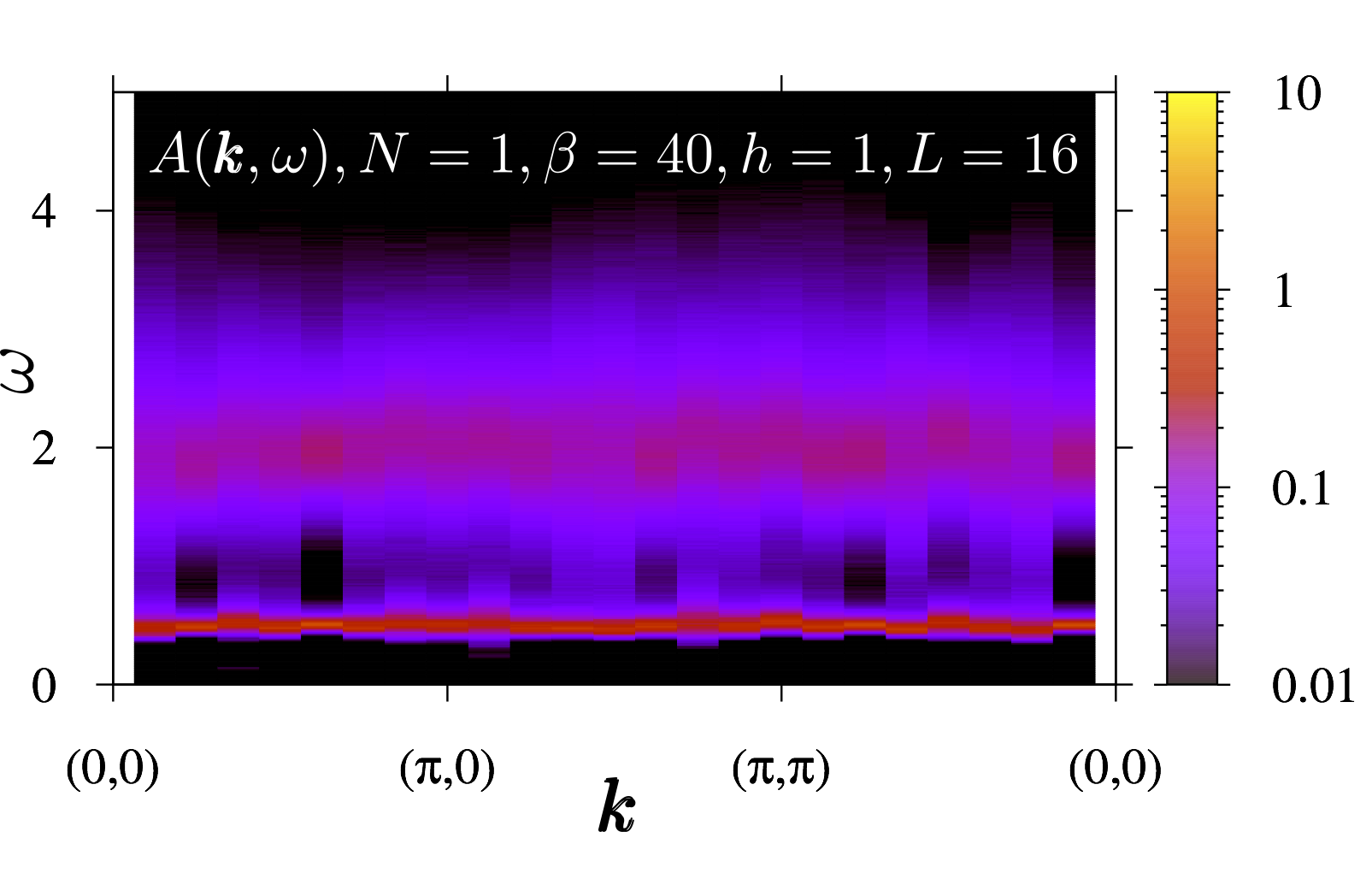}
\caption{ (color online)  Single-particle spectral function  $ A(\pmb{k},\omega) $ for the $N=1$ model.   As argued in the text, $A(\pmb{k},\omega)$ is $\pmb{k}$-independent. }
\label{QMC_test.fig}
\end{figure}

In the Majorana representation of Eq.~\ref{ModelMaj.Eq}   the O(2$N$) symmetry of our model  is manifest.  As a consequence, sign free quantum Monte Carlo (QMC) simulations can be carried out  for all values of $N$ \cite{Yao14a}.   We have used a  standard implementation of the finite temperature auxiliary field approach \cite{Hirsch81, Blankenbecler81,Assaad08_rev} to evaluate
 \begin{equation}
 	\langle \hat{O} \rangle   = \frac{\text {Tr} \left[  e^{-\beta \hat{H}  }  \hat{O} \right] }{\text {Tr} \left[  e^{-\beta \hat{H}  }  \right]}
 \end{equation}
 where the  trace runs over the fermionic and Ising degrees of freedom and $\beta$ corresponds to the  inverse temperature.  For Hubbard  type model simulations, the auxiliary field corresponds to a Hubbard-Stratonovitch field   with no explicit imaginary time dynamics \cite{Hirsch81, Blankenbecler81,Assaad08_rev}.  Here, in contrast, the auxiliary field corresponds to the z-component of the Ising field,   and  its imaginary time dynamics  is controlled by the transverse field. We have adopted a single spin flip algorithm, which turns out to be  efficient (high acceptance rates) at large values of $h$   where the field oscillates  rapidly in imaginary time. At weak couplings,  the Ising field becomes classical and  the single spin-flip update in space-time becomes increasingly inefficient. This renders simulations at low values of $h$ expensive.   Unless mentioned otherwise we have used  a Trotter time step  $\Delta \tau = 0.2$  so as to allow for a bond decomposition of  the infinitesimal time propagator:  $ e^{-\Delta \tau \hat{H} }  \simeq \prod_{\pmb{b}}  e^{ -\Delta \tau \hat{H}_{\pmb{b}} }  $, where  $\hat{H}_{\pmb{b}} $ corresponds to the Hamiltonian on a single bond.   Within the auxiliary field QMC method one can  compute  equal time as well as imaginary time displaced correlation functions.  We have carried out the analytical continuation with a stochastic implementation of the Maximum Entropy method \cite{Sandvik98,Beach04a}.   Eq.~\ref{Green_loc.eq}, which states that the single-particle Green's function is independent of momentum, provides an ideal test for the QMC as well as for the  analytical continuation since the locality of the single particle Green function results from the Monte Carlo  sampling.   Fig.~\ref{QMC_test.fig} shows the single particle spectral function $A(\pmb{k},\omega)  = -  \text {Im } G^{\text{ret}}(\pmb{k}, \omega) $ for $N=1$ model which is obtained from Wick rotation of the imaginary time Green function $G(\pmb{k},\tau) = - \langle T \hat{c}_{\pmb{k}}(\tau)    \hat{c}_{\pmb{k}}^{\dagger}(0) \rangle $. The $\pmb{k}$-independence of the data should be seen as a measure of  our spectral  resolution.
\section{Numerical results}

In this section we will describe our numerical results obtained from finite temperature auxiliary field simulations.   

\subsubsection*{\underline{\textbf{Summary of Results}}}

A sketch of the zero temperature phase diagram in the transverse field $h$ versus $N$ plane is shown in Fig.~\ref{Phase.fig}.   We will argue below that for all values of $N$, there is a finite temperature  Ising  transition at which the $\hat{Q}_{\pmb{i}} $ operators order. Since  $\left[ \hat{Q}_{\pmb{i}}, \hat{H} \right] =0$, this is akin to spontaneous symmetry breaking in a classical Ising model. For even (odd)   values of $N$  we  observe (anti) ferromagnetic  ordering of the 
$\hat{Q}_{\pmb{i}}$'s so that at zero temperature we can understand our results in terms of an effective $\mathbb{Z}_2$ gauge theory  coupled to 2$N$ Dirac fermions, with the understanding that non-gauge invariant unequal time, equal-space correlators \textit{do not} vanish, as discussed earlier. The (anti) ferromagnetic  ordering for even (odd)   values of $N$ is consistent with the fact that perturbatively, the couplings $K_{i,j}$ in Eq.~\ref{Ising.eq} for nearest neighbour vertices are proportional to $(-1)^{N+1}$. 

Spontaneous breaking of $\hat{Q}_{\pmb{i}} \rightarrow -\hat{Q}_{\pmb{i}} $  generates term proportional to $\sum_i  e^{i N \pmb{K}\cdot \pmb{i}} \hat{p}_{\pmb{i}}$ in the effective Hamiltonian, leading to spontaneous charge ordering of fermions below the finite temperature transition. The relevance/irrelevance of this term determines the fate of $h=0$ gapless Dirac fermions (Fig.~\ref{Sqom_h0.fig}) at infinitesimal $h$. When $N=1$, this term is a fermion bilinear proportional to $\sum_i  e^{i \pmb{K}\cdot \pmb{i}} \left(1 - 2\hat{n}_{\pmb{i}} \right)$ and thus leads to spontaneous charge-ordering and mass gap for fermions at infinitesimal $h$. For $N=2$, this term is  proportional to the Hubbard term $\sum_i \left(1 - 2\hat{n}_{\pmb{i},1} \right)\left(1 - 2\hat{n}_{\pmb{i},2} \right)$ where the overall sign of this term is determined by whether  $\langle \hat{Q}_{\pmb{i}}\rangle =1 $ or $-1$. The Hubbard term is irrelevant for gapless Dirac fermions, and therefore one expects to find a stable $\mathbb{Z}_2$ Dirac deconfined phase (Z$_2$D in Fig.\ref{Phase.fig}). Similarly, when $N=3$, the corresponding term $\sum_i  e^{i \pmb{K}\cdot \pmb{i}}  \left(1 - 2\hat{n}_{\pmb{i},1} \right)\left(1 - 2\hat{n}_{\pmb{i},2} \right) \left(1 - 2\hat{n}_{\pmb{i},3} \right)$ is irrelevant at the $h=0$ point, leading again to a stable Z$_2$D phase.

As $h$ is increased, the aforementioned Dirac deconfined phase undergoes a transition to a symmetry broken confined phase whose nature depends on the value of $N$ (Fig.\ref{Phase.fig}). For $N=2$, we find a N\'eel antiferromagnet, or a superconductor depending on whether $\langle \hat{Q}_{\pmb{i}}\rangle =-1 $ or $1$, respectively. For $N=3$, we find two symmetry broken phases, a dimerized ($\equiv$ Valence Bond Solid) phase separated from the Z$_2$D phase again by a seemingly continuous transition, and a N\'eel phase at larger values of $h$. The transition between the  spin-dimerized state and the N\'eel phase is apparently also continuous  and thus provides an example of deconfined quantum critical point \cite{Senthil04_2}.

\subsection{N=1}
As already discussed briefly in Sec.\ref{sec:model}, in   the large-$h$  limit, our model Hamiltonian (Eq.\ref{Model.Eq}) maps  onto (Eq.~\ref{h_inf.eq}) 
\begin{equation}
	\hat{H}_{\infty}  =       \frac{1}{ h} \sum_{ \left< \pmb{i},\pmb{j} \right> }   \left(  \hat{n}_{\pmb{i}} - 1/2\right)  \left(  \hat{n}_{\pmb{j}} - 1/2\right) 
\end{equation}
so that the ground state is  a charge density wave. In this limit  $ \langle \hat{X}_{\pmb{b} }\rangle  \simeq   1 $   and  the charge ordering  follows a $\pmb{K} = (\pi, \pi) $ modulation.    Consequently the $\hat{Q}_{\pmb{i}}$'s, by definition (Eq.\ref{eq:Q}), order anti-ferromagnetically.    Therefore, as a function of temperature, there must be a finite temperature transition in the 2D Ising universality class below which the $\hat{Q}_{\pmb{i}}$ develop long range order.   In the absence of a particle-hole symmetry breaking field,
$\langle \hat{Q}_{\pmb{i}}  \rangle = 0$  on any finite lattice and one  has to  measure  correlation functions  to detect ordering. The  $   \langle \hat{Q}_{\pmb{i}} \hat{Q}_{\pmb{j}} \rangle $  correlation functions turn out to be very noisy and  we have found it more convenient to include a small symmetry breaking field: 
\begin{equation}
\label{Stag_field.eq}
	\hat{H}_s   = h_s \sum_{\pmb{i}}  e^{i \pmb{K}\cdot \pmb{i}}  \left( \hat{n}_{\pmb{i}}  - 1/2  \right).
\end{equation}
Such a term in Hamiltonian does not introduce a negative sign problem  \cite{Li16} and pins  the CDW.   Since $ \left[  \hat{Q}_{\pmb{i}},  \hat{H}_s \right]  = 0 $,  $  \hat{Q}_{\pmb{i}}$  remains a good quantum number at finite value of $h_s$.  In fact  $\hat{H}_s$ acts on the Ising model of the  $ \hat{Q}_{\pmb{i}} $ variables as a staggered magnetic field. 

\begin{figure}[h]
\includegraphics[width=\linewidth]{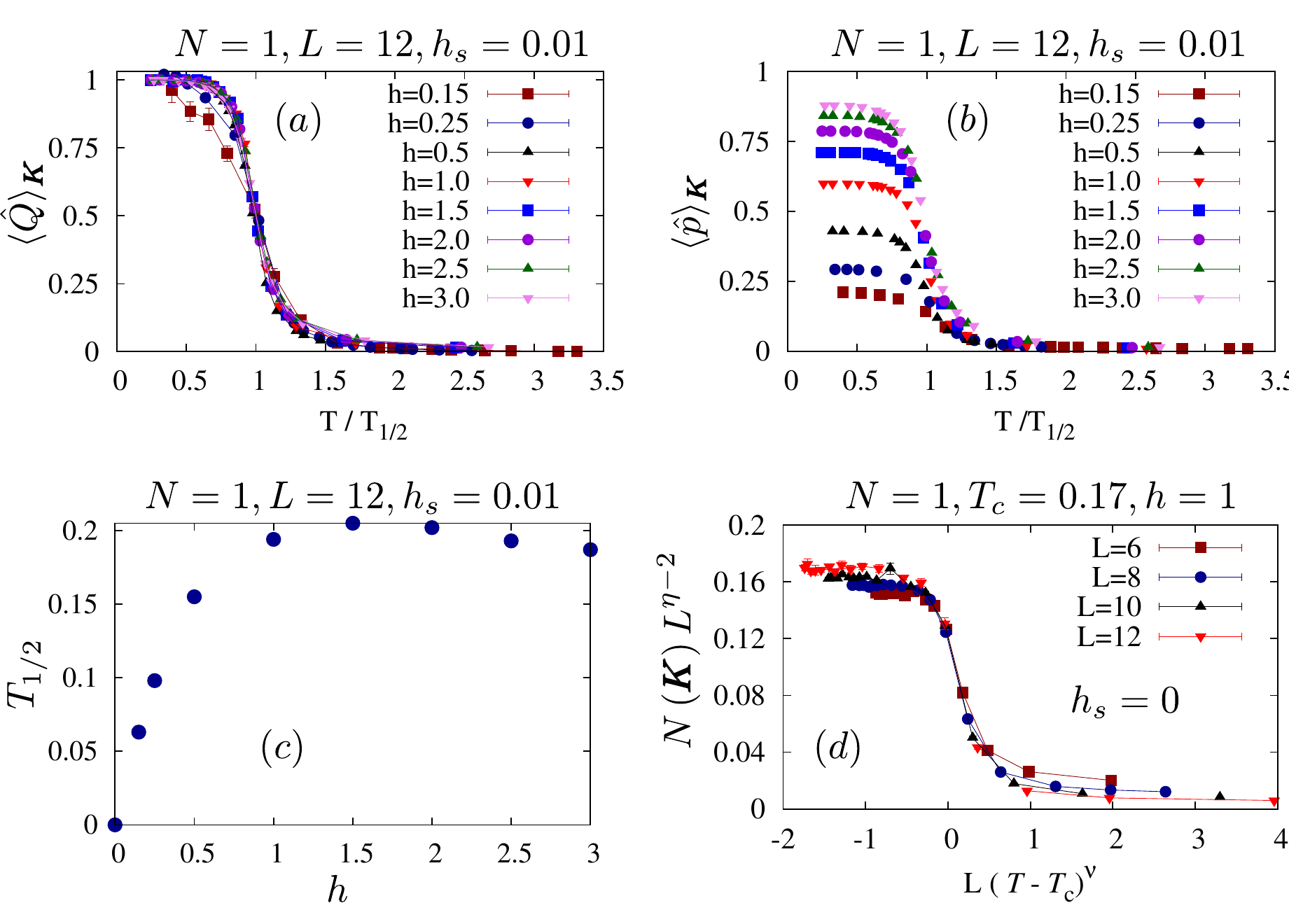} 
\caption{(color online)  (a)  Staggered order parameter of Eq.~\ref{U1_stag_order.eq} at finite  symmetry breaking field $h_s = 0.01$. 
(b)  Staggered order parameter of Eq.~\ref{U1_stag_order_ph.eq} again at  finite  symmetry breaking field $h_s = 0.01$.  
 The temperature scale is normalized by 
$T_{1/2}$ defined in Eq.~\ref{U1_T12.eq}.  (c)   $T_{1/2} $ as a function of $h$. (d) 
Data collapse of the density-density correlation function defined in Eq.~\ref{U1_Nq.eq}. This correlation function is related to the CDW order parameter via Eq.~\ref{eq:CDW}.  A reasonable data collapse is obtained using the 2D Ising exponents.}
\label{Tc_U1.fig}
\end{figure}
Figure \ref{Tc_U1.fig}(a)  plots  
\begin{equation}
\label{U1_stag_order.eq}
\langle \hat{Q} \rangle_{\pmb{K}} = \frac{1}{L^2}\sum_{\pmb{i}=1}^{L^2}    e^{i \pmb{K} \cdot \pmb{i}} \langle \hat{Q}_{\pmb{i}}  \rangle 
\end{equation} 
at a small value of the particle-hole symmetry breaking field, $h_s = 0.01 $. At low temperatures we find that $\langle \hat{Q} \rangle_{\pmb{K}} $ scales to unity  \textit{for all values of $h$}.  Since $\hat{Q}_{\pmb{i}} $ is a good quantum number, even in the presence of the symmetry breaking field, this means that the ground state has
\begin{equation}
	\hat{Q}_{\pmb{i}}  = e^{i \pmb{K} \cdot \pmb{i}}
\end{equation}
and the gauge constraint is dynamically generated.    The temperature scale at which  $ \langle \hat{Q} \rangle_{\pmb{K}} $  takes  half of its maximal value  defines 
$T_{1/2}$
\begin{equation}
\label{U1_T12.eq}
  \langle \hat{Q} \rangle_{\pmb{K}}  (T = T_{1/2} )  = 1/2.
\end{equation}
Figure~\ref{Tc_U1.fig}(c)  plots $T_{1/2}$ as a function of $h$.   The plot is consistent with the expectation that  $T_{1/2}$  vanishes at $h=0 $ and $h=\infty$ since in both  limits all  $\hat{Q}_{\pmb{i}} $  configurations are degenerate.    

Ordering of $\hat{Q}_{\pmb{i}}$ implies charge ordering. This is because the low-energy effective Hamiltonian now contains a term proportional to $ \sum_{\pmb{i}}  e^{i \pmb{K}\cdot \pmb{i}}  \hat{Q}_{\pmb{i}}  =   \sum_{\pmb{i}}  e^{i \pmb{K}\cdot \pmb{i}}  \hat{X}_{\pmb{i},\pmb{i} + \pmb{a}_x} \hat{X}_{\pmb{i},\pmb{i} - \pmb{a}_x} \hat{X}_{\pmb{i},\pmb{i} + \pmb{a}_y} \hat{X}_{\pmb{i},\pmb{i} - \pmb{a}_y} \hat{p}_{\pmb{i}}  $. At the same time, 
by symmetry, the effective Hamiltonian already contains a term proportional to $\sum_{\pmb{i}} \hat{X}_{\pmb{i},\pmb{i} + \pmb{a}_x} \hat{X}_{\pmb{i},\pmb{i} - \pmb{a}_x} \hat{X}_{\pmb{i},\pmb{i} + \pmb{a}_y} \hat{X}_{\pmb{i},\pmb{i} - \pmb{a}_y}$. The product of these two terms, which will also be generated, is proportional to   $\sum_{\pmb{i}} e^{i \pmb{K}\cdot \pmb{i}}  \hat{p}_{\pmb{i}} $  and leads to charge-ordering.  To confirm this, we have hence computed
\begin{equation}
\label{U1_stag_order_ph.eq}
\langle \hat{p} \rangle_{\pmb{K}} = \frac{1}{L^2}\sum_{\pmb{i}=1}^{L^2}    e^{i \pmb{K} \cdot \pmb{i}} \langle \hat{p}_{\pmb{i}}  \rangle. 
\end{equation} 
As apparent from Fig.~\ref{Tc_U1.fig}(b) the energy scale at which $\langle \hat{p} \rangle_{\pmb{K}}$  picks  up, matches $T_{1/2}$ defined in Eq.\ref{U1_T12.eq}.   As a further test 
 we have computed  charge-charge correlation functions in the absence of  symmetry breaking field: 
\begin{equation}	
\label{U1_Nq.eq}
	N(\pmb{q}) = \frac{1}{4}\sum_{\pmb{r}} e^{i \pmb{q} \cdot \pmb{r}} \langle \hat{p}_{\pmb{r}} \hat{p}_{\pmb{0}} \rangle 
\end{equation}
where  $\hat{p}_{\pmb{i}}    = \left(1 - 2\hat{n}_{\pmb{i}} \right) $. At low  temperatures, ordering  occurs at the antiferromagnetic wave vector $\pmb{K} = (\pi,\pi) $ and  as shown in Fig.~\ref{Tc_U1.fig}(c)  a reasonable data collapse is  obtained using the 2D Ising exponents, $\eta = 1/4$ and  $ \nu = 1$.    For the value of $h=1$ considered in Fig.~\ref{Tc_U1.fig}(d)   we obtain $T_c  \simeq 0.17$  which stands, to a first approximation, in agreement  with  the value of $T_{1/2}$  listed in Fig.~\ref{Tc_U1.fig}(b).  

\begin{figure}
\includegraphics[width=\linewidth]{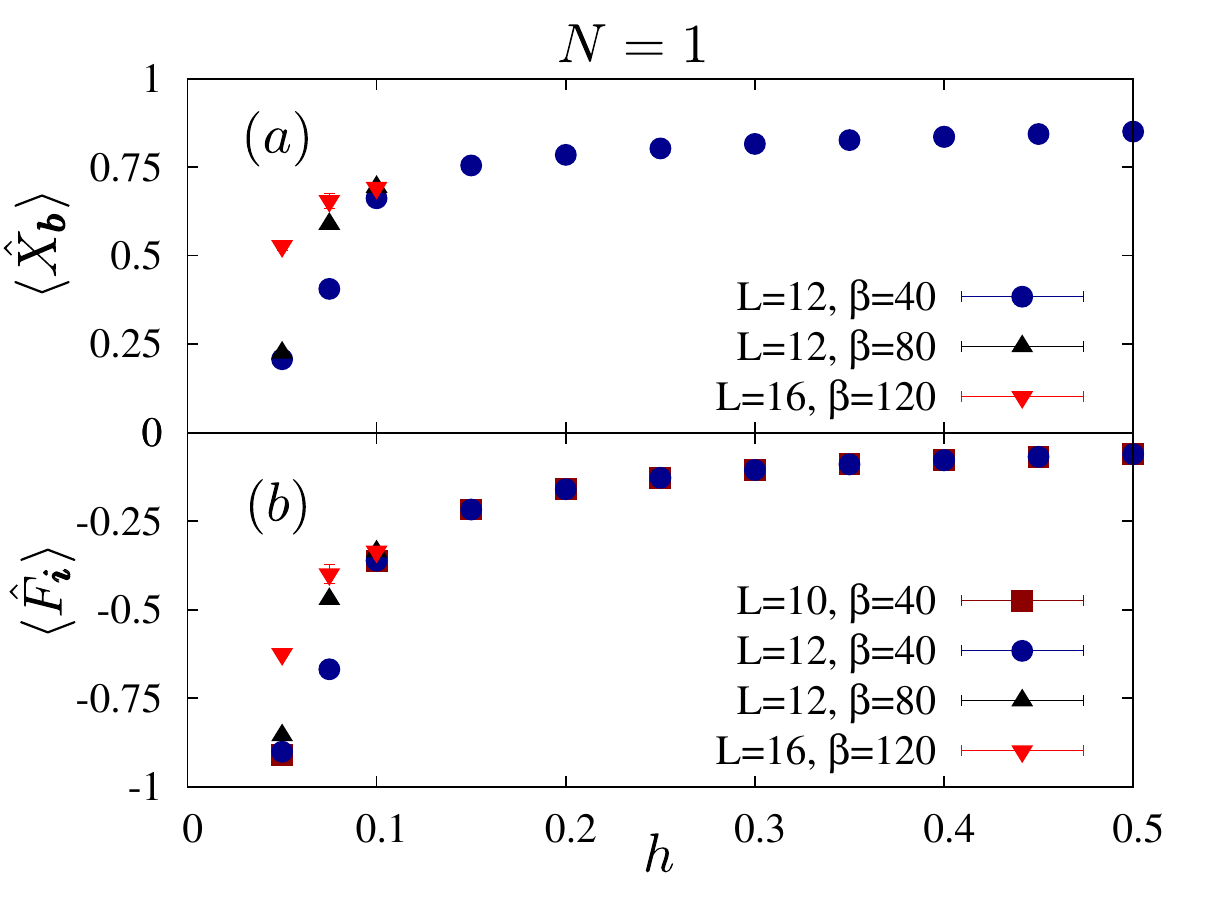}
\caption{ (color online)  (a) Electric field  and (b) magnetic flux  for the $N=1$ model as a function of the transverse field.  After taking into account the strong finite temperature and size effects in the small  $h$ limit,  the data is consistent with  a smooth behavior. }
\label{Ising_U1.fig}
\end{figure}
Fig.~\ref{Ising_U1.fig}   plots `gauge invariant' quantities  $\langle \hat{X}_{\pmb{b}} \rangle$  and $\langle \hat{F}_{\pmb{i}}   \rangle$  (Eq.\ref{Z2_flux.eq}) corresponding to the Ising field. We note that the  average electric field  is given by
\begin{equation}
\label{Electric.eq}
	 \langle \hat{X}_{\pmb{b}} \rangle =  \frac{1}{L^2} \frac{\partial F}{\partial h}
\end{equation}
where $F$ is the free energy, and therefore provides a direct access to the behavior of free energy as a function of $h$.
At   small values of $h$  one observes strong finite size  and temperature effects for both quantities.  Our lowest temperature results support a smooth crossover from  $ \langle \hat{X}_{\pmb{b}} \rangle  =0$, $ \langle \hat{F}_{\pmb{i}}   \rangle=-1$ to  $ \langle \hat{X}_{\pmb{b}} \rangle  =1$, $ \langle \hat{F}_{\pmb{i}}   \rangle=0$  as $h$ grows.     This implies that there is no level crossing  associated with a change of ground state  quantum numbers and that the gauge constraint remains  enforced for all values of the transverse field.    At $h=0$  all  plaquettes are  threaded by a  $\pi$-flux static magnetic field and no electric field  is present.  Starting form this limit  gauge fluctuations introduce pairs of $\mathbb{Z}_2$ vortices  -- which are nothing but plaquettes with zero flux --  such that as a function of $h$   the magnetic field grows.    The fluctuations of the magnetic field induce the onset of the electric field. 

\begin{figure}
\includegraphics[width=\linewidth]{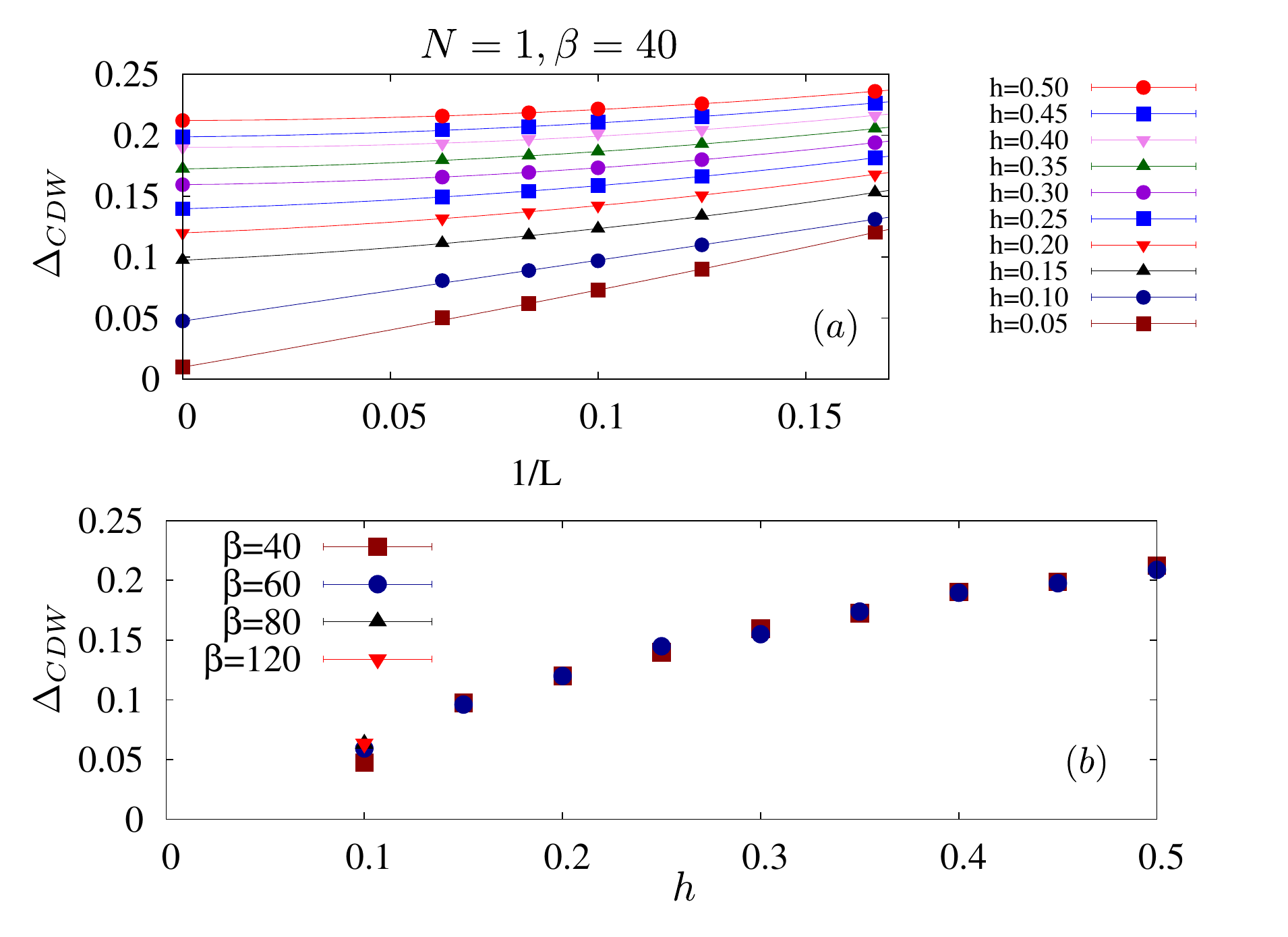}
\caption{(color online)  (a) Size extrapolation of  the charge density wavve order parameter $\Delta_{CDW}$ performed  with a second order polynomial in $1/L$.  (b) Extrapolated value of  $\Delta_{CDW}$ at different temperatures.  Below  $h = 0.1 $ our lattice sizes are too small to unambiguously extrapolate  to the thermodynamic limit. }
\label{CDW_U1.fig}
\end{figure}

We now discuss zero-temperature phase diagram. The  CDW order parameter, 
\begin{equation}
	 \Delta_{CDW} = \sqrt{N(\pmb{K})/L^2 }, \label{eq:CDW}
\end{equation} 
at low temperatures and as a function of $h$ is plotted in Fig.~\ref{CDW_U1.fig}.  Consistent with the fact that at $h = 0$, the spectrum is gapless (Fig.~\ref{Sqom_h0.fig}), and correspondingly $T_{1/2} \rightarrow 0$ as $h \rightarrow 0$ (Fig.~\ref{Tc_U1.fig}), the CDW order parameter also approaches zero as $h \rightarrow 0$. We expect that $\langle \hat{Q}_{\pmb{i}} \rangle \neq 0$ for any non-zero $h$, and therefore the Z$_2$D phase is stable only at $h=0$.

\begin{figure}
\includegraphics[width=0.9\linewidth]{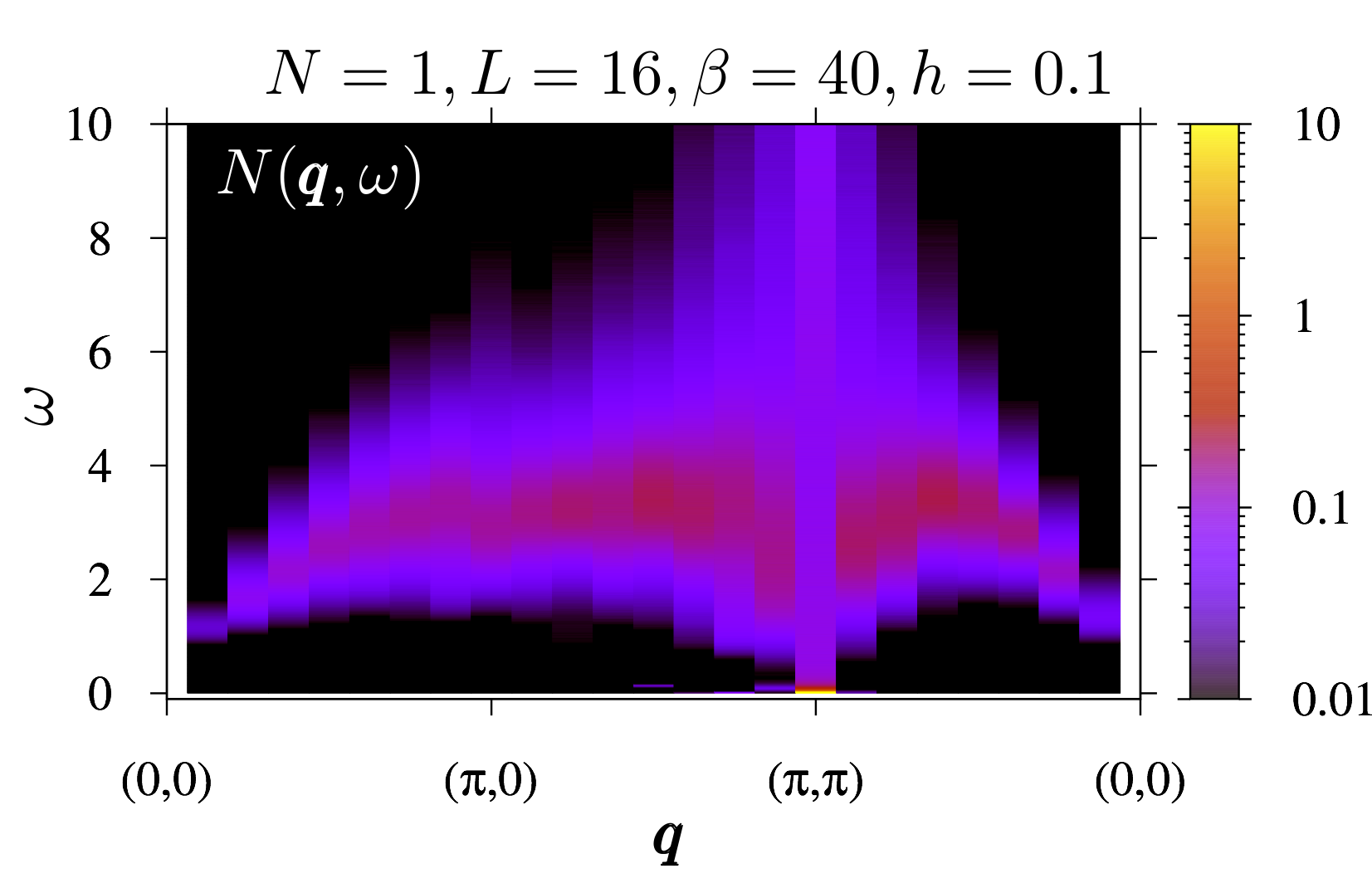}
\caption{ (color online)  Dynamical charge structure factor for the $N=1$ model.  }
\label{Dynamcis_U1.fig}
\end{figure}

In Fig.~\ref{Dynamcis_U1.fig}   we  consider  the  dynamical properties of the model at $h = 0.1 $ and $\beta = 40$.   This choice of parameters places us below $T_{1/2}$  and we will   discuss the data   in terms of the ground state  with quantum numbers
\begin{equation}  
	\hat{Q}_{\pmb{i}} | \Psi_0 \rangle  =  e^{i \pmb{K} \cdot \pmb{i}}    | \Psi_0 \rangle. 
\end{equation}
 The dynamical  charge structure factor reads, 
\begin{equation}
	N(\pmb{q} , \omega)  = \pi \sum_{n}   |  \langle \Psi_n | \hat{N}_{\pmb{q}}   | \Psi_0 \rangle  |^2   \delta ( E^{N}_n - E^{N}_0 - \omega )  \label{spectral.eq}
\end{equation}
with 
\begin{equation}
	\hat{N}(\pmb{q})  =  \frac{1}{L}   \sum_{\pmb{i}}  e^{i \pmb{q}\cdot \pmb{i}} \hat{n}_{\pmb{i}}.
\end{equation}
The intermediate states, $ | \Psi_n \rangle $, have the same  $\hat{Q}_{\pmb{i}}$  quantum numbers  as  in the ground state  since $ \left[ \hat{n}_{\pmb{i}} , \hat{Q}_{\pmb{i}} \right]  = 0$.   Thereby, the dynamical charge structure factor of Fig.~\ref{Dynamcis_U1.fig} is identical  to  that  one would obtain when imposing  exactly the gauge constraint on the  Hilbert space.  Within this  constrained Hilbert space,  the  elementary  excitations  are closed Wilson loops of  $\hat{Z}_{\pmb{b}}$ operators in Euclidian space-time as well as open strings of $\hat{Z}_{\pmb{b}}$   operators with {\it particle-hole} excitations at the extremities again in $2+1$ dimensions. 
For the considered values of  $h$ the ground state  shows long range CDW order and  the gauge string binds {\it particle-hole}  excitations.  This corresponds to the confined phase of the gauge field and  leads to the  central peak feature  at the  wave vector $\pmb{K}=(\pi,\pi) $ apparent  in Fig.~\ref{Dynamcis_U1.fig}.      In the limit $h\rightarrow  \infty$ this becomes the dominant feature of the spectral function.  At higher energies  or on shorter time scales,  the string does not bind  {\it particle-hole} excitations and a continuum very reminiscent of the $h=0$ limit is apparent in the spectral function.

\subsection{N=2} \label{sec:N=2}

\begin{figure*}
\includegraphics[width=1.0\linewidth]{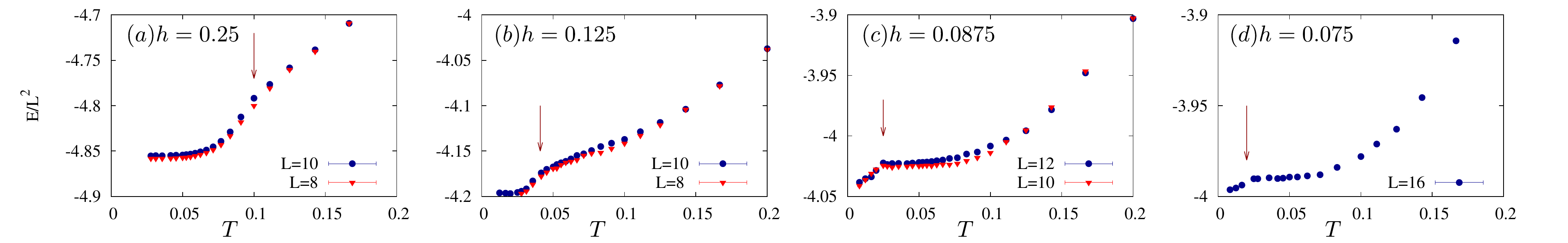} 
\caption{(color online)  Energy as a function of temperature for the $N=2$ model at various values of $h$.}
\label{Etot_SU2.fig}
\end{figure*}

As discussed in the Sec.\ref{sec:sym_cont}, for $N=2$ the global symmetry is O(4). To understand the physical content of this symmetry, it is instructive to consider the limit $h \rightarrow \infty$:

\begin{eqnarray}
	\hat{H}_{\infty}  =     & &   - \frac{1}{4 h} \sum_{ \left< \pmb{i},\pmb{j} \right>}   \left[\sum_{\sigma=1}^{2}   \left( \hat{c}^{\dagger}_{\pmb{i},\sigma} \hat{c}_{\pmb{j},\sigma}  + \text{H.c.}  \right) \right]^2 \nonumber \\  = & & 
	\frac{1}{h} \sum_{ \left< \pmb{i},\pmb{j} \right> }   \left[ \hat{\pmb{S}}_{\pmb{i} }   \hat{\pmb{S}}_{\pmb{j} }   + \hat{\pmb{\eta}}_{\pmb{i} }   \hat{\pmb{\eta}}_{\pmb{j} }  \right].
\end{eqnarray}
Here, $\hat{\pmb{S}}_{\pmb{i} } = \frac{1}{2}\sum_{\sigma, \sigma'} \hat{c}^{\dagger}_{\pmb{i},\sigma} \pmb{\sigma}_{\sigma,\sigma'} \hat{c}_{\pmb{i},\sigma'} $ corresponds to a fermionic representation of the spin operator  and $ \hat{\pmb{\eta}}_{\pmb{i} }  = \hat{P}^{-1}_{1}\hat{\pmb{S}}_{\pmb{i} } \hat{P}_{1} $ where $\hat{P}_{1}$  corresponds to particle-hole symmetry transformation defined in Eq.~\ref{eq:ph}.  Since the $\pmb{\eta}  $-operators are derived from the spin operators with a similarity transformation they satisfy the SU(2) algebra as well. Furthermore, $\left[ \hat{\eta}_{\pmb{j},\alpha},  \hat{S}_{\pmb{i},\beta} \right] = 0 $ so that at the level of Lie algebra, the O(4) symmetry can be understood as  O(4) $\cong$ SU${}_{S}(2)$ $\otimes$ SU${}_{\eta} (2) $ \cite{Yang90}, \footnote{At the level of group structure, O(4) $\cong \frac{SU_{S}(2) \otimes SU_{\eta} (2)}{\mathbb{Z}_2} \otimes \mathcal{J}$ where $\mathcal{J}$ implements improper rotation, but this distinction will not be relevant for us}.

The  Ising order parameter corresponding to particle-hole symmetry  breaking (Eq.~\ref{eq:parity}) relates to the $\hat{\pmb{S}}$- and $\hat{\pmb{\eta}}$-operators via 
\begin{equation}
	\hat{p}_{\pmb{i}}  = \frac{4}{3} \left(    \hat{\pmb{\eta}}_{\pmb{i}}^2  -  \hat{\pmb{S}}_{\pmb{i}}^2\right).  
\end{equation} 
If particle-hole symmetry breaks, then there are two possibilities. If  $ \langle \hat{p}_{\pmb{i}} \rangle  < 0 $, then the SU${}_{S}(2)$ $\otimes$ SU${}_{\eta} (2)$ is reduced to SU${}_{S}(2)$. In the extreme limit of $p=-1$, sites are singly occupied, and the model reduces to a spin-only model. Away from this limit, there are charge fluctuations per site, but  the universal properties (such as the phase transition out of the deconfined phase), which depends only on symmetries, can still be understood in terms of only spin fluctuations. The other possibility is that $ \langle \hat{p}_{\pmb{i}} \rangle  > 0 $, in which case the sites are predominantly empty or doubly occupied, leading to a charge-only model in the extreme limit $p=1$. In this case, the underlying symmetry is  SU${}_{\eta} (2)$ and the phase diagram can be understood in terms of fluctuations of a three component vector consisting  of CDW (one component) and s-wave superconductor ( two components, since it has both real and imaginary components).  Since the superconducting phase, whenever it exists, is always degenerate with CDW due to SU${}_\eta(2)$ symmetry, henceforth we will  call the corresponding phase as just superconductor.

\begin{figure}
\includegraphics[width=\linewidth]{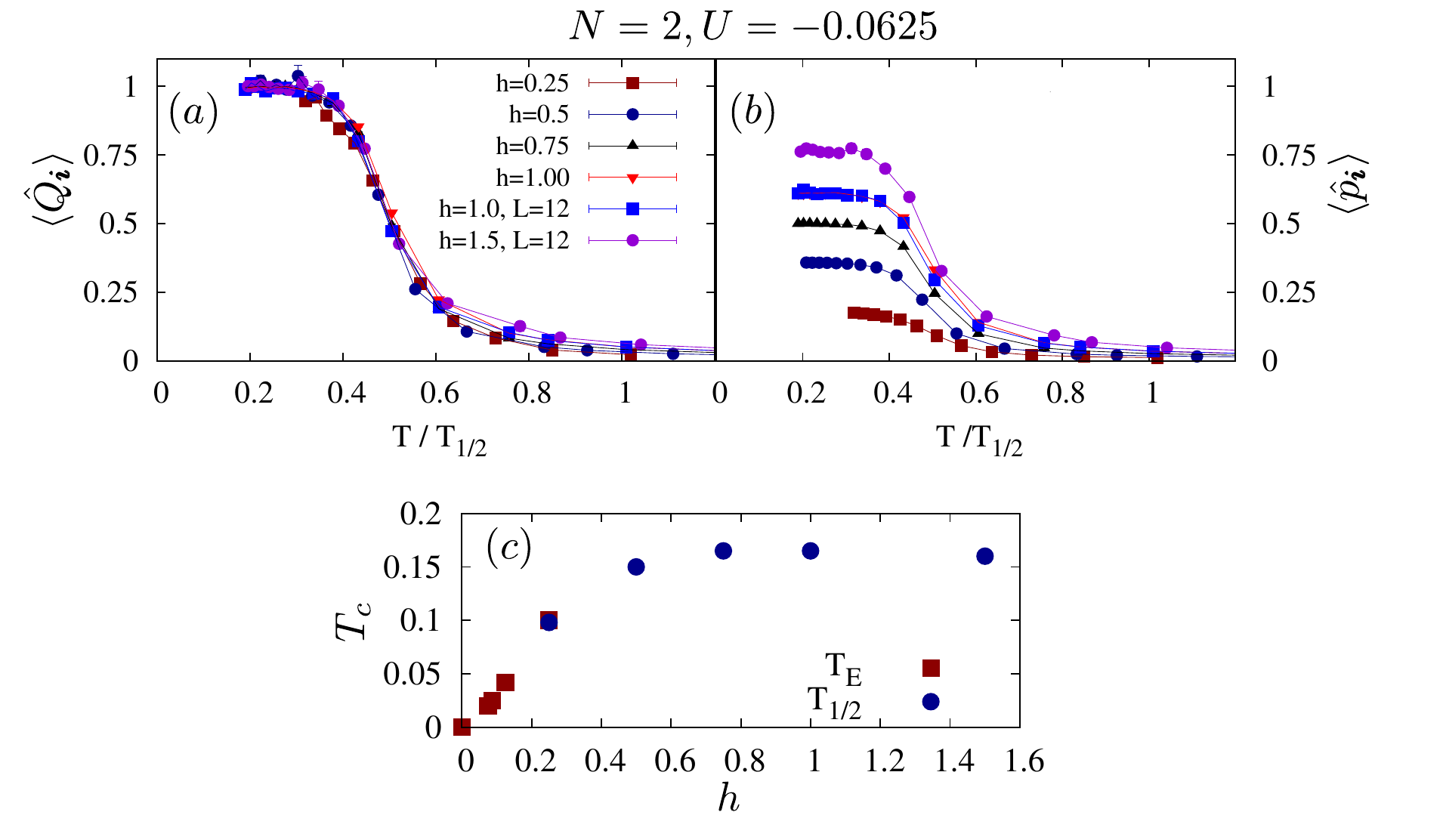}

\caption{(color online)  (a) Average value of $\langle \hat{Q}_{\pmb{i}} \rangle $  in the presence of a {\it small}  symmetry breaking field, $U = -0.0625$. 
(b) Average value of $\langle \hat{p}_{\pmb{i}} \rangle $ at the same value of the symmetry breaking field.  Unless mentioned otherwise, we have  set $L=8$. (c)   Estimate of $T_c$ below which we expect   $\langle \hat{Q}_{\pmb{i}} \rangle $ to order ferromagnetically. $T_E$ stems from  the energy data of Fig.~\ref{Etot_SU2.fig}   and the temperature scale $T_{1/2}$ is  defined by  $\langle \hat{Q}_{\pmb{i} }\rangle(T_{1/2}) =1/2$.  }
\label{Q_SU2_scan.fig}
\end{figure}
To detect the possibility of Ising transitions corresponding to ordering of  $\hat{Q}_{\pmb{i}}$ and $\hat{p}_{\pmb{i}}$ we compute the   temperature dependence of the internal energy for a range of values of $h$ (Fig.~\ref{Etot_SU2.fig}).  Note that in contrast to the $\hat{Q}_{\pmb{i}}$'s,  $ \hat{p}_{\pmb{i}}$ is not a good quantum number since
$\left[ \hat{p}_{\pmb{i}}, \hat{H} \right] \neq 0 $. The Ising transition in two-dimensions has a  specific heat exponent $\alpha = 0$  which implies that the specific heat is logarithmic  divergent at the transition temperature.   This results in an infinite slope of the internal energy   at $T_c$ and in the  thermodynamic  limit.  An arrow in each plot  points to the temperature  where singularities are apparent. It is beyond the scope of our calculations to pin down the  details of the singularity.  In particular, when $T_c$ is {\it small}  very large systems sizes  are required to resolve the two dimensional nature of the singularity.   Within our  resolution we observe only one transition which we will now argue corresponds to the ordering of the $ \hat{Q}_{\pmb{i}}$'s as well as  $\hat{p}_{\pmb{i}}$, that is, breaking of particle-hole symmetry. 

\begin{figure}
\includegraphics[width=0.9\linewidth]{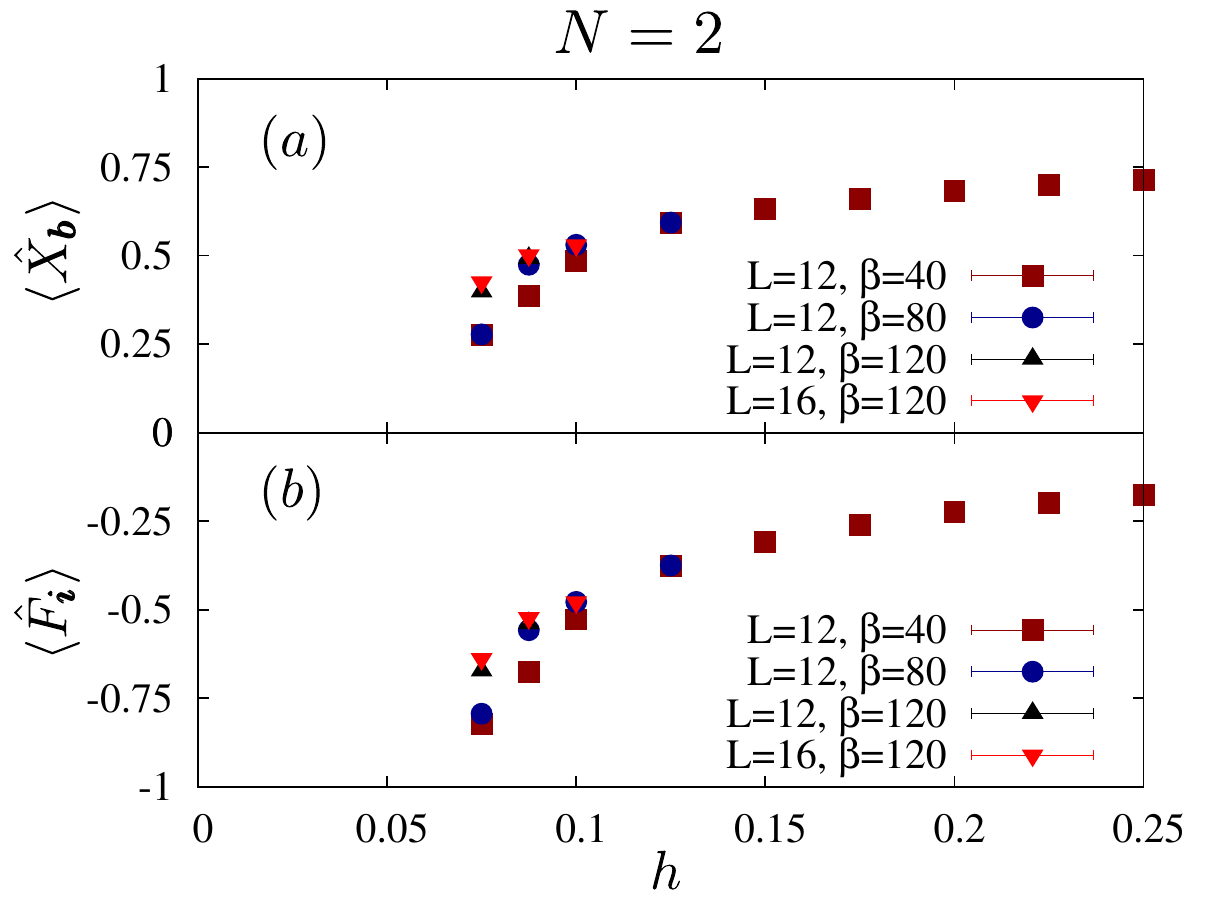} 
\caption{(color online)   (a) Electric field  and  (b) magnetic flux per plaquette  as a function of $h$ for the $N=2$ model.   }
\label{Flux_SU2.fig}
\end{figure}

To explicitly see the ordering of  $ \hat{Q}_{\pmb{i}}$ and  $\hat{p}_{\pmb{i}}$, we  add a {\it small}  attractive Hubbard term to the model, $\hat{H}_U = -U \sum_{\pmb{i}} \left(  \hat{n}_{\pmb{i},1} -1/2 \right) \left(  \hat{n}_{\pmb{i},2} -1/2 \right)$. This does not introduce a sign problem \cite{Wu04},  but breaks particle-hole symmetry and effectively acts as a uniform  magnetic field for the $  \hat{Q}_{\pmb{i}} $ variables, alongside the Ising interaction term of Eq.~\ref{Ising.eq}.  Note that   $ \left[  \hat{H}_U,\hat{Q}_{\pmb{i}}\right] =0 $ so that  $\hat{Q}_{\pmb{i}}$ remains a good quantum number upon inclusion of this symmetry breaking term. Fig.~\ref{Q_SU2_scan.fig}  shows $ \langle \hat{Q}_{\pmb{i}} \rangle $ as a function of temperature at small values  of the Hubbard $U$.  At an energy scale set by $T_c$ one observes a growth in $ \langle \hat{Q}_{\pmb{i}} \rangle $   which saturates to  the value $ \langle \hat{Q}_{\pmb{i}} \rangle \simeq 1 $.  The fact that $ \langle \hat{Q}_{\pmb{i}} \rangle \simeq 1 $   at low  temperature implies ferromagnetic  coupling between the $ \hat{Q}_{\pmb{i}} $ variables in Eq.~\ref{Ising.eq}.   Since the Hubbard term  acts as a Zeeman field for particle-hole symmetry as well,  we can compute $ \langle \hat{p}_{\pmb{i}} \rangle $  to check if particle-hole symmetry is broken spontaneously at the same energy scale.     If so, one expects  $\langle \hat{p}_{\pmb{i}} \rangle $  to show a marked growth at the energy scale   where   $\langle \hat{Q}_{\pmb{i}} \rangle  \approx 0.5$.  Fig.~\ref{Q_SU2_scan.fig}(b)   supports this. 
Fig.~\ref{Q_SU2_scan.fig}(c)  summarizes our estimate of $T_c$ as obtained from  the internal energy and  by  explicitly including a symmetry breaking term.

Below  $T_c$, any change  in the $ \hat{Q}_{\pmb{i}} $ quantum numbers   as a function of $h$   should  correspond to a level crossing.  To rule out this scenario, we  plot  the  average electric (see Eq.~\ref{Electric.eq})  and magnetic field per plaquette (Eq.~\ref{Z2_flux.eq}) as a function of $h$.   Size and temperature effects  are strong in the weak coupling limit where $T_c$ is small. However, our lowest temperature results on  the $L=16$ lattice are consistent with a smooth evolution  of  the magnetic flux  and electric field.  Hence, at an  energy scale set by $T_c$ the constraint is dynamically generated  such that  at $T=0$ we can interpret our findings in terms of a $\mathbb{Z}_2$  lattice gauge theory with  $\hat{Q}_{\pmb{i}} = 1$, or $\hat{Q}_{\pmb{i}} =-1$ coupled to fermions.

As discussed above, the pattern of ordering for  $ \hat{Q}_{\pmb{i}} $  determines the pattern of symmetry breaking of the continuous global symmetries in the confined phase. Specifically, superconducting phase corresponds to $\hat{Q}_{\pmb{i}} = 1$ and  $\langle \hat{p}_{\pmb{i}} \rangle > 0 $ while the SDW phase corresponds to $\hat{Q}_{\pmb{i}} = -1$ and  $\langle \hat{p}_{\pmb{i}} \rangle < 0 $. One can in fact explicitly map a superconducting state to an SDW state by particle-hole transformation. Consider a superconducting ground state $|\Psi\rangle$, so that  

\begin{equation}
	\left(\hat{H} + \hat{H}_U \right) |\Psi \rangle =  h | \Psi \rangle  \text{   and  }   \hat{Q}_{\pmb{i}} |\Psi \rangle = | \Psi \rangle.
\end{equation}  
 
Since $\left\{ \hat{P}_{\uparrow},  \hat{H}_U  \right\} =0 $ and $\left\{ \hat{P}_{\uparrow} , \hat{Q}_{\pmb{i}} \right\} =0 $,  it follows that one can generate an orthogonal state 
$|\Psi' \rangle   =  \hat{P}_{\uparrow} |\Psi \rangle $ which satisfies
\begin{equation}
	\left( \hat{H} - \hat{H}_U \right) |\Psi' \rangle =  h | \Psi '\rangle  \text{   and  }   \hat{Q}_{\pmb{i}} |\Psi' \rangle = -| \Psi' \rangle.  
\end{equation} 
If the state $|\Psi\rangle$ corresponds to a superconductor, then $|\Psi'\rangle$ corresponds to SDW, since $\langle \Psi| \hat{\pmb{S}}_{\pmb{i} } .  \hat{\pmb{S}}_{\pmb{j} }|\Psi\rangle = \langle \Psi'| \hat{\pmb{\eta}}_{\pmb{i} }.   \hat{\pmb{\eta}}_{\pmb{j}}|\Psi'\rangle$ for all ${\pmb{i} }, {\pmb{j} }$. In the limit $U \rightarrow 0$, $|\Psi'\rangle$ would also correspond to an eigenstate of the Hamiltonian. As an aside, as discussed in the introduction, when the number of sites is odd, then $|\Psi\rangle$ and $|\Psi'\rangle$ are  supersymmetric partners \cite{Hsieh16}, and thus $\mathcal{N} = 2$ supersymmetry relates superconducting and SDW order parameters. Similar dichotomy between superconductor and SDW was also explored in \cite{Feldbach03}.

\begin{figure}
\includegraphics[width=\linewidth]{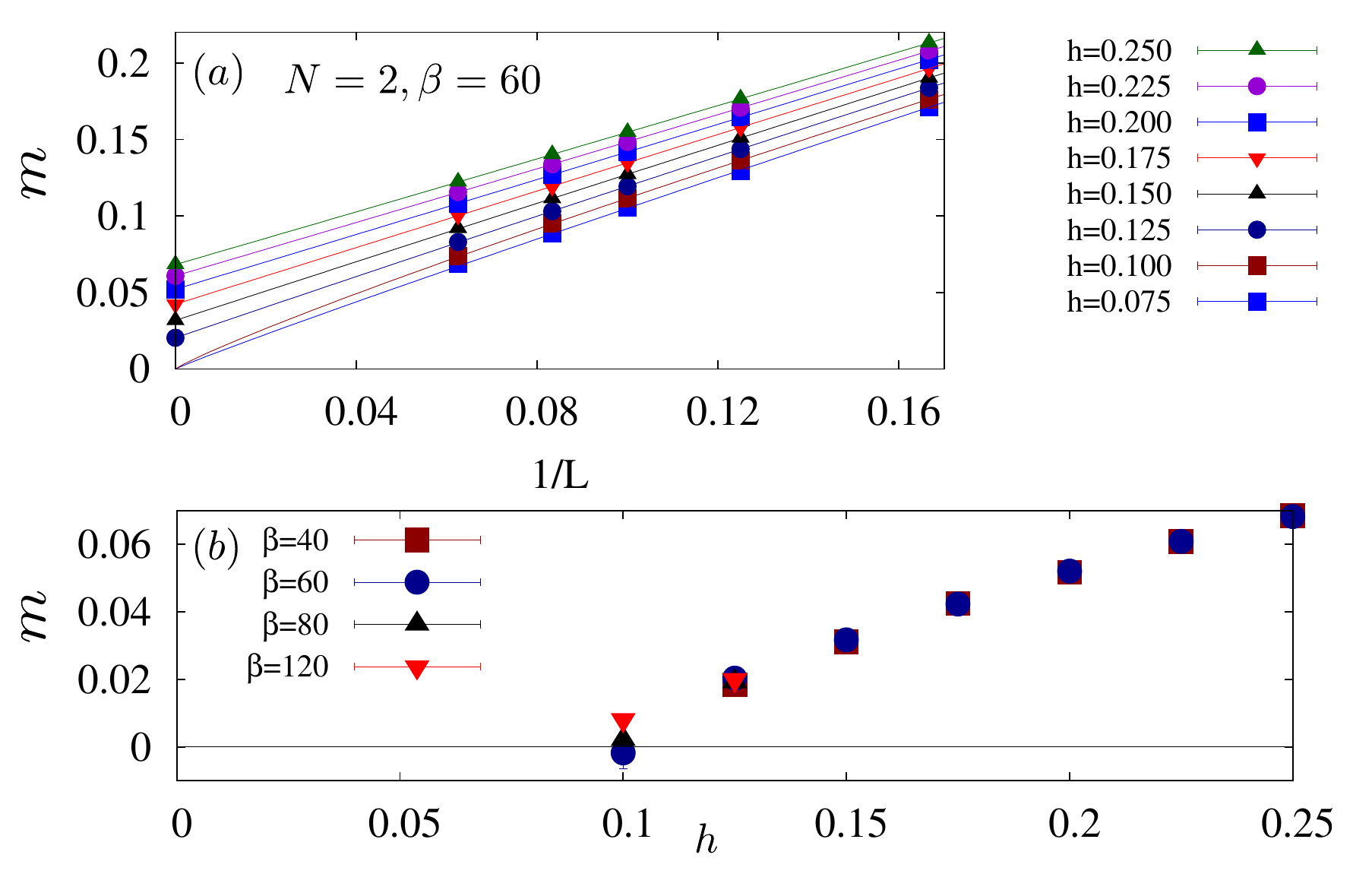}
\caption{(color online)   (a) Magnetization as a function of system size (see Eq.~\ref{Magnetization.eq}).   In the ordered phase  we have used a second order  polynomial form in $1/L$  to extrapolate to the thermodynamic limit.   At $h= 0.1$ and $h= 0.075$ we  have set $\beta  = 120$ and fitted the data to the  form $f(x) = aL^{-b}$. 
(b) Estimated value of the magnetization in the thermodynamic limit.  
  }
\label{Spin_SU2.fig}
\end{figure}

\begin{figure*}
\includegraphics[width=1.0\linewidth]{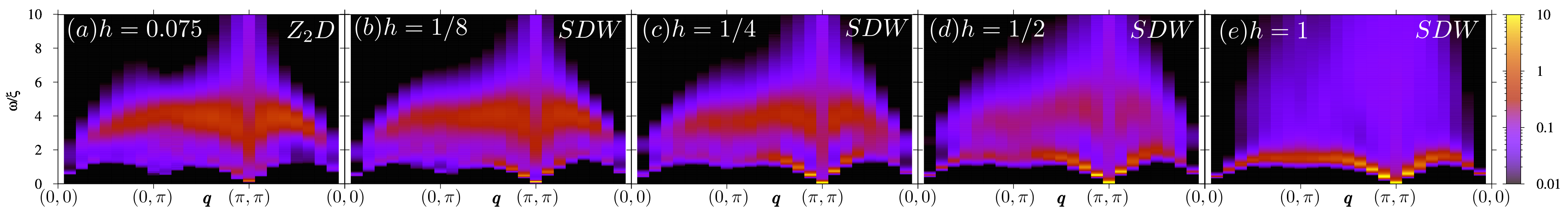} 
\caption{(color online)  Dynamical spin structure factor for the $N=2$ model  as a function of $h$.  For  the smallest value of $h$ we have considered $\beta = 60$  as opposed to 
$ \beta  = 40$   for other values of the transverse field.  This  choice of temperature places us below $T_c$ such that $\hat{Q}_{\pmb{i}} \simeq 1 $ }
\label{Spin_Dynamics_SU2.fig}
\end{figure*}

\begin{figure}
\includegraphics[width=\linewidth]{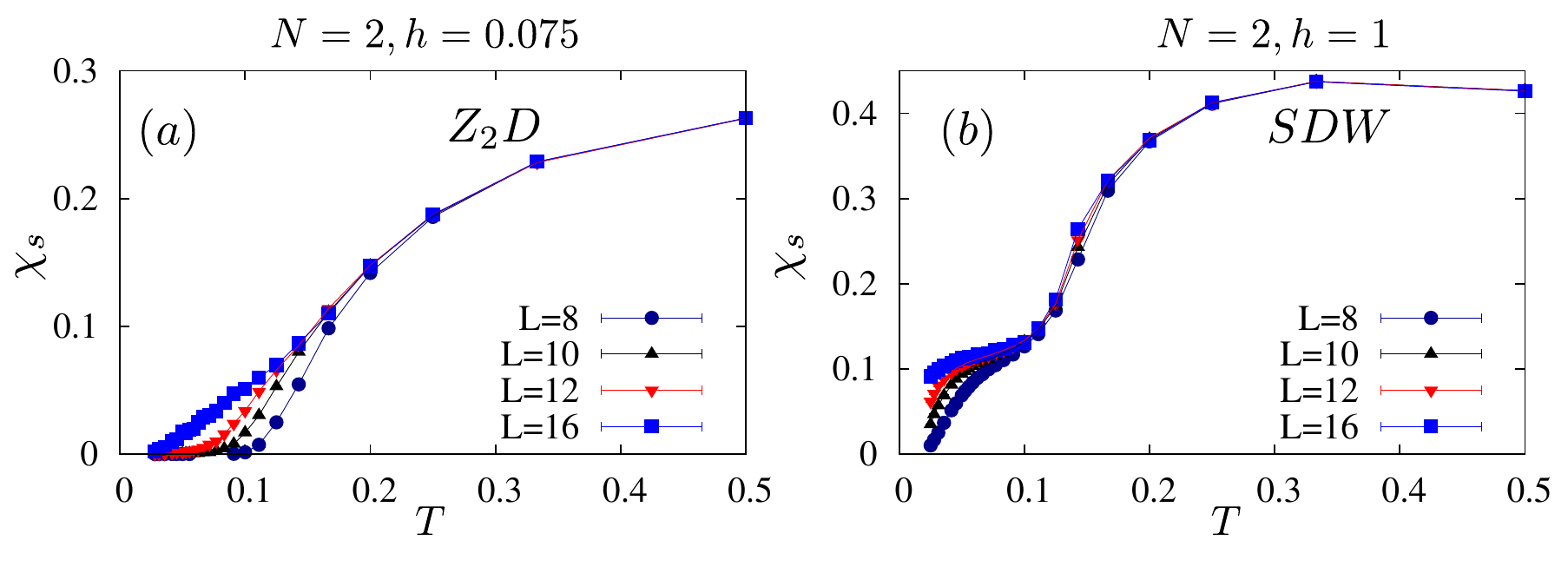}
\caption{(color online)  Uniform spin susceptibility  at a)  $h=0.075$ (Z$_2$D phase) and  b) $h=1$  (SDW phase). }
\label{Spin_Sucep_SU2.fig}
\end{figure}

Let us  investigate the zero temperature in the language of spins i.e. in the sector $ \hat{Q}_{\pmb{i}} = -1 $. We compute  the spin-spin correlations given by: 
\begin{equation}
	S_{\text{Spin}}(\pmb{q},\tau) =  \frac{1}{L^2} \sum_{\pmb{r},\alpha,\beta} e^{i \pmb{q}\cdot \pmb{r}} \langle  \hat{S}^{\alpha}_{ \; \; \beta}(\pmb{r}, \tau)   \hat{S}^{\beta}_{ \; \; \alpha}(\pmb{0}, 0) \rangle. 
\end{equation}
The  local moment of the antiferromagnetic order   is then defined as:
\begin{equation}
\label{Magnetization.eq}
	m = \sqrt{\frac{1}{L^2} S_{\text{Spin}}(\pmb{K},0) }.
\end{equation} 
Note that   $ \left[ \hat{S}^{\alpha}_{ \; \; \beta}(\pmb{i}),   \hat{Q}_{\pmb{i}} \right]=0 $ such that spin excitations  do not violate the dynamically generated constraint  at $T=0$.  Fig.~\ref{Spin_SU2.fig} shows the finite size scaling of $m$ for various value of $h$. We see that extrapolation to the thermodynamic limit with a second order form in $1/L$ yields a magnetization which vanishes  at $h \simeq 0.1 $.   Below   $h \simeq 0.1 $  we have tested for bond dimerized orders  but found no positive  signal. In this regime, we will  interpret our findings  in terms of a $\mathbb{Z}_2$-Dirac deconfined phase of our  effective lattice gauge theory.  At $T=0$   where $\hat{Q}_{\pmb{i}} = -1$,  the gauge invariant elementary excitations  are pairs of spinons tied by a gauge field string of $\hat{Z}_{\pmb{b}} $ operators  as well as closed loops of $\hat{Z}_{\pmb{b}} $'s.   Below $h \simeq 0.1 $   the gauge field is deconfined  and the magnetic flux per plaquette is close to $\pi$. In  particular one expects the  dynamical spin structure factor to bear  similarities with the particle-hole  continuum  
of Dirac fermions shown in Fig.~\ref{Sqom_h0.fig}.    In fact the Fig.~\ref{Sqom_h0.fig} would correspond to the mean-field theory of the $\mathbb{Z}_2$-Dirac deconfined phase.

 Fig.~\ref{Spin_Dynamics_SU2.fig} plots the dynamical spin-spin correlation functions as a function of $h$.   As apparent, in the small $h$ limit we indeed see a continuum of excitations  with soft features at wave vectors $(0,\pi)$ and $(\pi,\pi)$  corresponding to  transitions between Dirac nodes.   As $h$ grows spectral weight accumulates around $ \pmb{K} = (\pi,\pi) $  and a distinct spin-density wave feature emerges.   As in the $N=1$ case,  in the proximity of the transition the continuum of fractionalized excitations  is visible at high energies  whereas the low energy  sector is dominated by  the Goldstone mode. 

To see the difference between the small $h$ vs large $h$ phase, we  also considered  the spin-susceptibility, 
\begin{equation}
	 \chi_s = \frac{\beta}{N} \left( \langle \hat{\pmb{S}}^2  \rangle -  \langle \hat{\pmb{S}}  \rangle^2 \right).
\end{equation}
Some care has to be taken when considering this quantity, since Monte Carlo averages over different $ \hat{Q}_{\pmb{i} } $  sectors, and it  is only in the low temperature limit that   this quantity will correspond to that of the lattice gauge theory.  Our results are shown in Fig.~\ref{Spin_Sucep_SU2.fig}.    As apparent, the two phases have very distinct low energy signatures.  Taking into account finite size effects, the  $\mathbb{Z}_2$-Dirac deconfined  phase is characterized by a $\chi_s \propto T $  in accord with the mean-field theory of this state. On the other hand,  the confined phase   is consistent with a  constant low temperature spin-susceptibility  originating from spin-wave excitations.

\subsection{N=3}

\begin{figure}
\includegraphics[width=\linewidth]{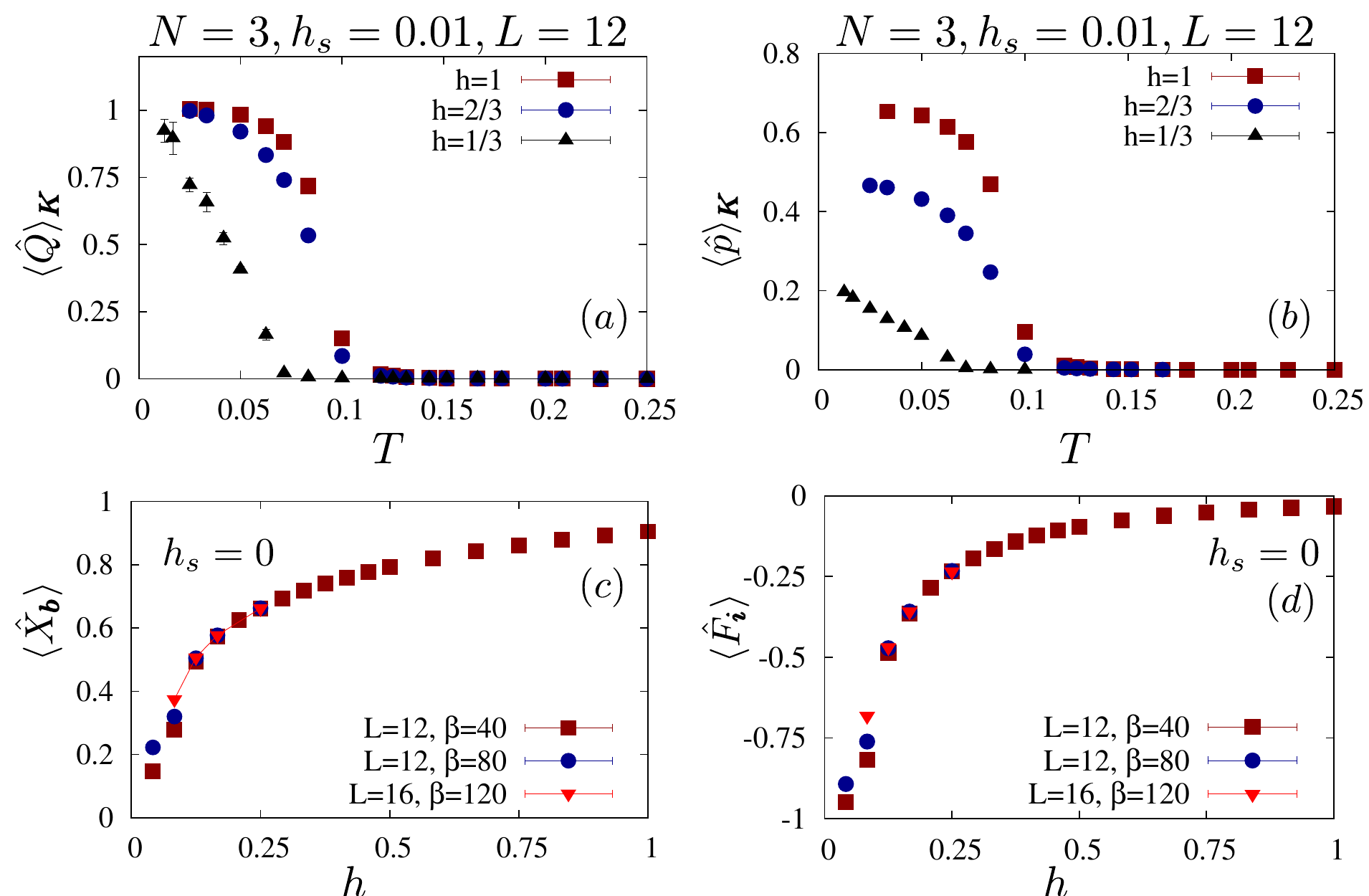}
\caption{(color online)  (a)   $\langle \hat{Q} \rangle_{\pmb{K}} = \frac{1}{L^2}\sum_{\pmb{i}} e^{i \pmb{K} \cdot \pmb{i}} \langle \hat{Q}_{\pmb{i}} \rangle $   and 
  (b) $\langle \hat{p} \rangle_{\pmb{K}} = \frac{1}{L^2}\sum_{\pmb{i}} e^{i \pmb{K} \cdot \pmb{i}} \langle \hat{p}_{\pmb{i}} \rangle $ in the presence of a small staggered chemical potential term which breaks particle-hole symmetry and introduces an  antiferromagnetic Zeeman term for the Ising Hamiltonian of Eq.~\ref{Ising.eq}.  
(c) Electric field and (d)   Magnetic flux per plaquette  as a function of $h$.  }
\label{Q_SU3_scan.fig}
\end{figure}

We will next  discuss the $N=3$ model (Eq.\ref{Model.Eq}) where alongside a deconfined phase at small $h$ and  magnetically ordered phase at large $h$,  a \textit{spin dimerized} state emerges at intermediate values of $h$.  
As in the $N = 1$ case, the $\hat{Q}_{\pmb{i}}$'s  order anti-ferromagnetically such that we can pick up the ordering temperature by  implementing  a small staggered field as in Eq.~\ref{Stag_field.eq}.
Our results are plotted in Fig.~\ref{Q_SU3_scan.fig}(a) and (b).  As expected,  anti-ferromagneitc ordering of  $\hat{Q}_{\pmb{i}}$  and  $\hat{p}_{\pmb{i}}$ occurs at the same energy scale. Being a constant of motion  even in the presence of the small staggered chemical potential  $\hat{Q}_{\pmb{K}}$  (see Eq.~\ref{U1_stag_order.eq}) saturates to unity.  
As a technical aside, we  note for values of $h < 1/3$   the Monte Carlo  autocorrelation time  for aforementioned quantities  becomes  large  thus rendering the calculation difficult. 

In Fig.~\ref{Q_SU3_scan.fig}(c) and (d)    we plot the magnetic flux as well as the electric field as a function of $h$.    Size and temperature  effects are large at   a small values of $h$. Nevertheless,  our data is consistent with absence of level crossing  in the  ground state quantum numbers since  owing to Eq.~\ref{Electric.eq} this would result in a jump  in the electric field $\langle \hat{X}_{\pmb{b}}  \rangle $. 

\begin{figure}
 \includegraphics[width=\linewidth]{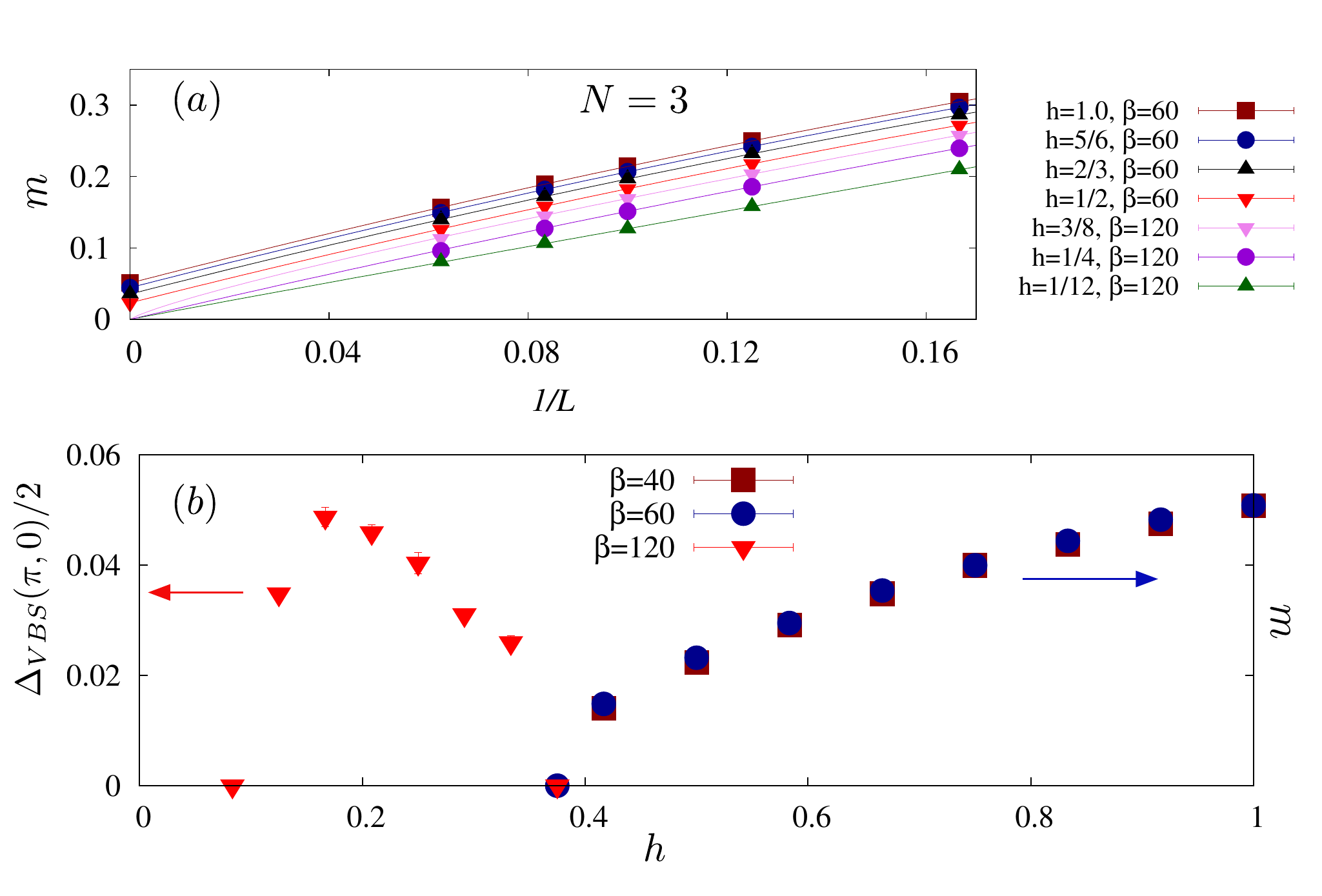}
 \caption{(color online)  (a)  SU(3) spin staggered moment as a function of system size.    The data is consistent with the following choices of fit functions. For $h > 3/8$ we have  used a second order polynomial in 1/L.  At $h=3/8$ we have fitted the data to the form    $m(L)     = aL^{-(1+\eta)/2}  $ to obtain $\eta \simeq 0.64$  
  which  turns out to be substantially  larger than the exponent  obtained for the  J-Q model for the SU(3) case, $\eta \simeq  0.4$ \cite{Block13}.   For lower values of $h$ we again use a polynomial fit and impose  a zero intercept with the y-axis.   In  Fig.~(b) we plot the extrapolated value of the magnetization as well as of the VBS order parameter. At $h=3/8$ both VBS and SDW order parameters are consistent with a power-law decay.  } 
\label{Stag_mag_SU3.fig}
\end{figure}

The anti-ferromagnetic ordering of $\hat{p}_{\pmb{i}}$  along with the constraint that there are on average  one and a half electrons per unit cell,  gives us two options for the resulting charge ordering: \\ I. One can  place two charges on sub-lattice A and one charge on sub-lattice B.  This choice leaves  the SU(3) color degree of freedom unquenched. In particular, on sub-lattice A, two SU(3)  spins combine to give $3 \otimes 3 = \overline{3} \oplus 6 $ where 3 corresponds to the fundamental representation,    $\overline{3} $ to the antisymmetric complex conjugate representation with Young  tableau  consisting of  a single column and  two rows and  6  to the symmetric representation with  Young tableau consisting of a single row and two columns. Since we are working with fermions, only the  $\overline{3}$  representation is relevant and  has states given by
\begin{equation}
	\Delta^{\dagger}_{\alpha,\pmb{i}}   = \frac{1}{2} \epsilon_{\alpha,\beta,\gamma} \hat{c}^{\dagger}_{\beta,\pmb{i}}\hat{c}^{\dagger}_{\gamma,\pmb{i}}.
\end{equation}
Here summation over repeated indices,  running over  the three colors of SU(3), is implied.  Neglecting the site index $\pmb{i}$ and assuming that under an infinitesimal SU(3)  rotation  $ \hat{c}^{\dagger}_{\gamma} \rightarrow    \left( 1 + i \epsilon \pmb{e} \cdot \pmb{T} \right)_{\gamma, \delta} \hat{c}^{\dagger}_{\delta} $ with $ \pmb{T} $,  the vector of  Gell-Mann matrices,  $\pmb{e}$ a unit vector in $\mathbb{R}^8$, and $\epsilon$ a real infinitesimal, one will show that $\Delta^{\dagger}_{\alpha} \rightarrow \left( 1 - i  \epsilon \pmb{e} \cdot \overline{ \pmb{T}} \right)_{\alpha, \delta} \Delta^{\dagger}_{\delta} $.   On sub-lattice B  the single electron corresponds   to the fundamental representation of SU(3).  With  this assignment of representations,   N\'eel  states, as well  spin-dimerized states can  naturally occur.      For the  N\'eel state  we break the SU(3) symmetry by arbitrarily choosing a color $\alpha$  so as to define the state: 
\begin{equation}
\label{NeelSU3.eq}
  | \text{N\'eel}  \rangle  = \prod_{\pmb{i} \in \text{A}} \Delta^{\dagger}_{\alpha,\pmb{i}}   \prod_{\pmb{j} \in \text{B}} \hat{c}^{\dagger}_{\pmb{j},\alpha} | 0 \rangle
\end{equation} 
Fluctuations around this state will produce spin waves.   On the other hand,  a possible  VBS state  with $\pmb{q}   = (\pi,0) $  order is given by:
\begin{equation}
	| \text{VBS}  \rangle  = \prod_{ \pmb{i},   e^{i \pmb{q}\cdot \pmb{i}}=1}   \left(   \sum_{\alpha=1}^{3} \Delta^{\dagger}_{\alpha,\pmb{i}} c^{\dagger}_{\alpha,\pmb{i}+\pmb{a}_x} \right) |0\rangle.   
\end{equation}
This state remains invariant under SU(3) rotations and thereby defines a singlet. \\
 II. The other charge ordering pattern consistent with the anti-ferromagnetic structure of the $\hat{p}_{\pmb{i}} $'s, corresponds  to three particles on  sub-lattice B and none on sub-lattice A. This assignment quenches the spin degree of freedom.

 One might expect that  SDW states which possess low lying Goldstone modes are energetically favorable. Our QMC results are consistent with this expectation. In Fig.\ref{Stag_mag_SU3.fig}  we plot the equal time spin-spin   correlation  and corresponding staggered moment $m$.   Antiferromagnetic spin ordering is present  down to $h \simeq 3/8$. \footnote{Note that under a particle-hole transformation, the SDW phase will map onto the color superconducting states discussed in Ref.~\cite{Honerkamp04a}. There is however an important  difference, namely the unpaired flavor forms an CDW rather that a Fermi liquid. }
At lower values of $h$,   the data is consistent with the vanishing of SDW.   
\begin{figure*}
\includegraphics[width=0.9\linewidth]{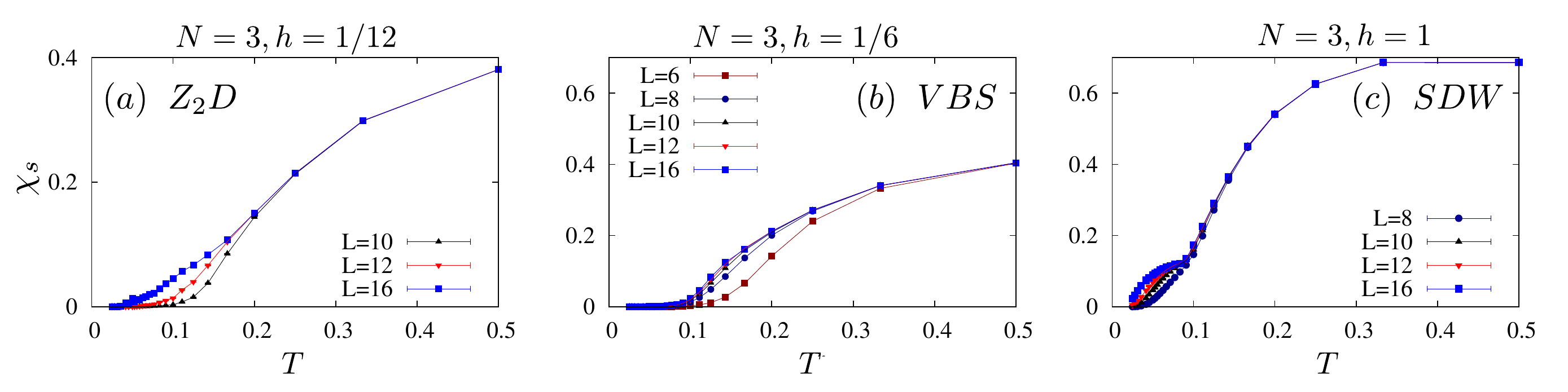}
\caption{(color online)  Uniform spin susceptibility  in the  Z$_2$D,  VBS and SDW phases. }
\label{Suscep_SU3.fig}
\end{figure*}

\begin{figure}
  \includegraphics[width=\linewidth]{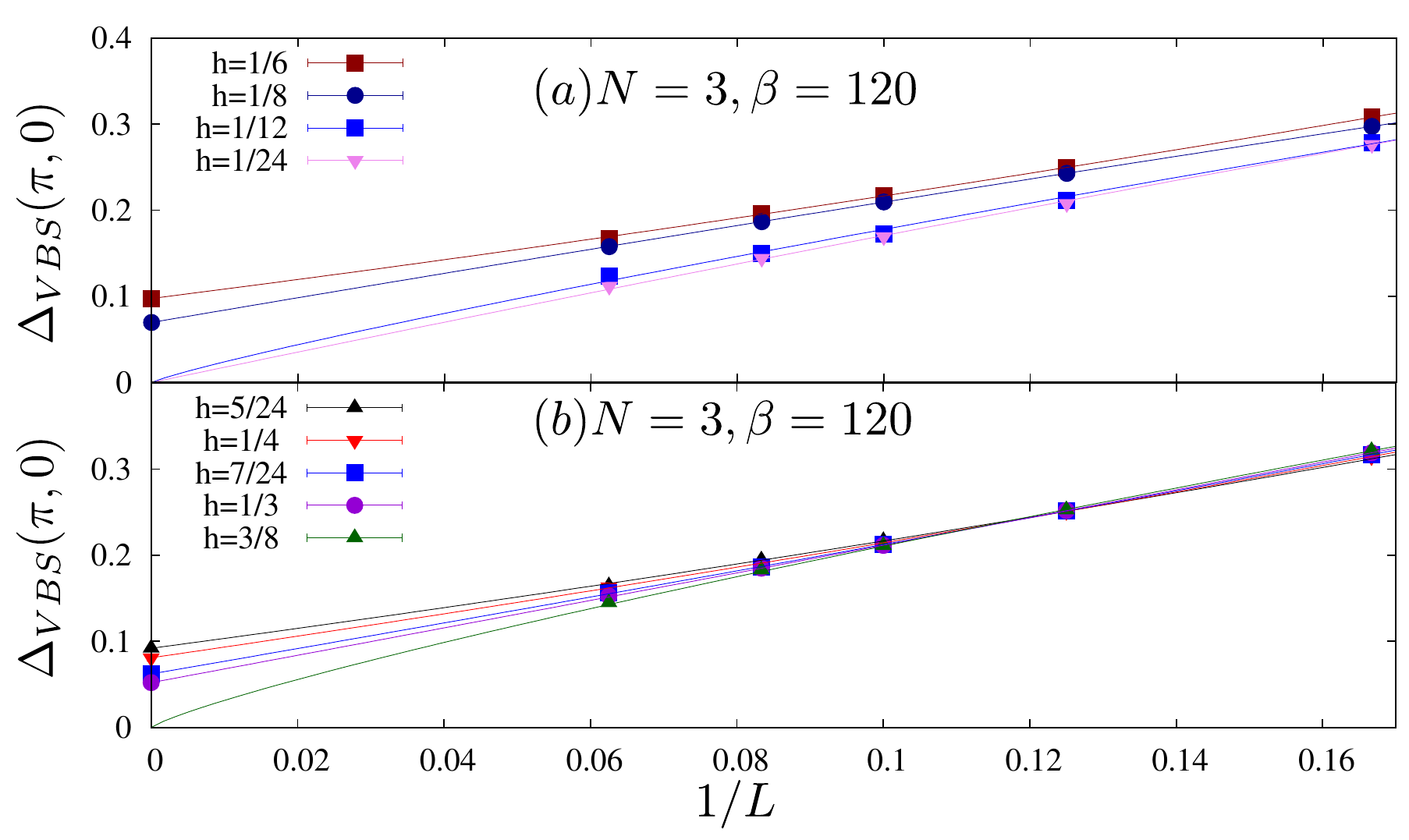}
 \caption{(color online)   SU(3)  VBS order parameter.  At $h = 1/12$   corresponding to a rough estimate of the transition   from the $\mathbb{Z}_2$ Dirac deconfined to the dimerized  phase,  we fit the data to the form: 
 $a L^{-b} $.  The transition from the VBS phase to the SDW occurs  approximately at $h = 2/3$. For this value of $h $  the data is consistent with a large $\eta = 2b -1$  exponent, $\eta \simeq 0.6$ which is  larger than the one obtained   for the J-Q model at SU(3)  $ \eta \simeq 0.5$ (see Ref.~\cite{Block13} ). }
\label{Dimer_SU3.fig}
\end{figure}

The uniform spin susceptibility plotted in Fig.~\ref{Suscep_SU3.fig}  gives a very clear  account on the  nature of the  phases one expects  as a function of $h$.     At $h=1$  and at low temperatures,  the data is consistent with  the  SDW state with constant density of states originating  from the Goldstone modes. At  $h=1/12$ we observe, similar to $N=2$ case at $h = 0.075$, the characteristic  spin susceptibility of a Dirac deconfined phase. There is however a distinct intermediate phase possessing no low lying spin excitations.   As hinted  above, a very natural candidate to account for this is spin-dimerization which breaks the $C_4$ lattice symmetry.  To verify this, we compute the dimer-dimer correlation functions: 
\begin{eqnarray}
	    \left[ S_{\text{VBS}}  (\pmb{q})\right]_{\pmb{\delta},\pmb{\delta}' }   = & &  \\
	     \frac{1}{L^2} \sum_{\pmb{i},\pmb{j}}  e^{i \pmb{q} \cdot \left( \pmb{i} - \pmb{j} \right) } & &  \left( 
	      \langle \hat{\Delta}_{\pmb{i},\pmb{i}+\pmb{\delta}} \hat{\Delta}_{\pmb{j},\pmb{j}+\pmb{\delta}'} \rangle  -  
	       \langle \hat{\Delta}_{\pmb{i},\pmb{i}+\pmb{\delta}} \rangle \langle \hat{\Delta}_{\pmb{j},\pmb{j}+\pmb{\delta}'} \rangle  \right) \nonumber
\end{eqnarray}
with
\begin{equation*}
     \hat{\Delta}_{\pmb{i},\pmb{i}+\pmb{\delta}}   =   \hat{S}^{\alpha}_{ \; \; \beta}(\pmb{i})   \hat{S}^{\beta}_{ \; \; \alpha}(\pmb{i} + \pmb{\delta}) 
\end{equation*}
\footnote{To compute this correlation function we have set: $ \hat{S}^{\alpha}_{ \; \; \beta}(\pmb{i})  = \hat{c}^{\dagger}_{\pmb{i},\alpha} \hat{c}^{}_{\pmb{i},\beta}    - \frac{1}{2}  \delta_{\alpha,\beta} $ }.
From the above correlation function, we define   the  dimer order parameter as:
\begin{equation}
     \Delta_{VBS}(\pmb{q})   = \sqrt{  \frac{1}{L^2} \text{Tr}  \left[ S_{\text{VBS}}  (\pmb{q}) \right] }.
\end{equation}
Fig.~\ref{Dimer_SU3.fig}  shows a finite size scaling of the above VBS order parameter at wave vector $\pmb{q} = \left( \pi, 0 \right) $.   Extrapolating to the thermodynamic limit necessarily implies a choice for the fit function. In the parameter range where we observe  no low lying spin-excitations in the susceptibility,  we  use a  polynomial fit up to second order in $1/L$.    This fit indeed confirms long range valence-bond order  in the vicinity of $h = 1/6$.

 \begin{figure*}
 \includegraphics[width=0.95\linewidth]{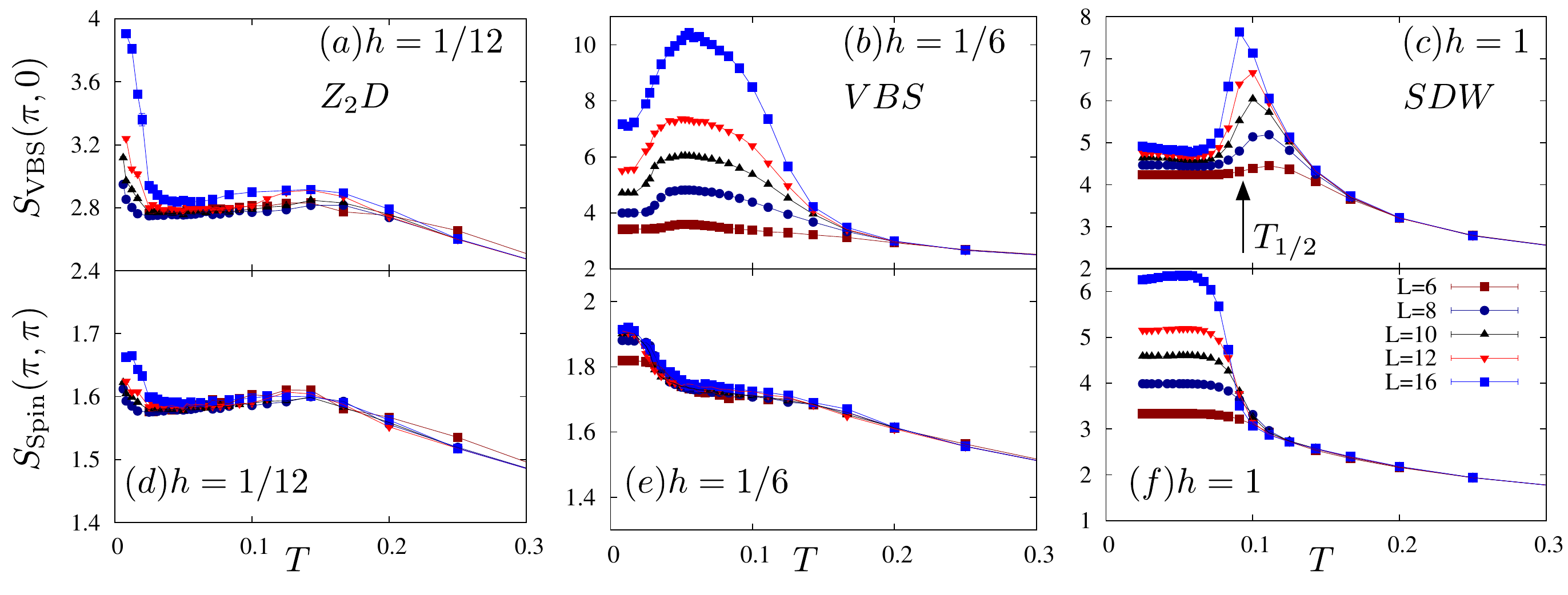} 
\caption{(color online) Temperature dependence of the dimer-dimer   and spin-spin  correlations in  the Z$_2$D, VBS and SDW phases of the $N=3$ model. The arrow in Fig.(c) corresponds to the energy scale at which $\langle \hat{Q}\rangle_{\pmb{K}}$  plotted in Fig.~\ref{Q_SU3_scan.fig} (a) takes a value of $1/2$ (see Eq.~\ref{U1_T12.eq}). }
\label{Tscan_SU3.fig}
\end{figure*}

In Fig.~\ref{Tscan_SU3.fig} we show the  temperature dependence of the  VBS and SDW  correlation functions at $\pmb{q}=(\pi,0)$ and at  $\pmb{q} = \left( \pi, \pi \right) $  respectively  and in the Z$_2$D, VBS and SDW phases.     The VBS order breaks the lattice $C_4$ symmetry and therefore one correspondingly expects a finite temperature   Kosterlitz-Thouless transition  below which $ S_{\text{VBS}}(\pi,0)$ diverges as a function of system size.   The SDW ordering breaks a continuous symmetry  and at  finite  temperatures the real space spin-spin correlations will decay exponentially if the total system size is bigger than the correlation length.  In our model, there is also the Ising transition of the  $\pmb{Q}_{\pmb{i}}$'s which  will   feed into the spin and dimer correlation functions.  
At $h=1$   we  can evaluate  $T_{1/2}$,  $\langle \hat{Q}\rangle_{\pmb{K}}(T_{1/2})=1/2$,  from the data shown in  Fig.~\ref{Q_SU3_scan.fig}(a).  This energy scales,  which tracks the Ising transition temperature, is  shown in Fig.~\ref{Tscan_SU3.fig} (c).  Here,  the VBS correlations show strong size effects above the $T_{1/2}$ scale  suggesting that  the Kosterlitz-Thouless transition temperature exceeds   the ordering temperature of the  $\pmb{Q}_{\pmb{i}}$'s.   At $T_{1/2}$ charge ordering equally occurs (see Fig.~\ref{Q_SU3_scan.fig}(b))   leading to  a {\it dynamical} selection of  the representations of  SU(3)  on each sub-lattice. The fact that at $T_{1/2}$  the antiferromagnetic spin correlations (Fig.~\ref{Tscan_SU3.fig} (f)) grow   agrees with the $3$ and $\bar{3}$  representation pattern.   The drop in the VBS order parameter at this energy scale  illustrates the  competition between these two ordered states and the data at $h=1$  supports  dominant   SDW correlations in the low temperature limit.  We  note that  the data of  Fig.~\ref{Tscan_SU3.fig} (f)  seemingly supports long range order at finite temperature.  However, recall that the antiferromagnetic correlation length grows exponentially with  inverse temperature  \cite{Chakravarty88} so that  when it exceeds our largest lattice size the QMC data  falsely suggests long range ordering.  Our results  at $h=1/6$  in the VBS phase are plotted in  Fig.~\ref{Tscan_SU3.fig} (b),(e). Comparing  with the h=1 data set, one observes very similar finite  temperature behavior and we again interpret the sharp downturn in the VBS correlation functions as a measure of the Ising transition. However, and  in contrast to the $h=1$ data set, the low energy  results  are consistent with  long range dimer order.  In the Z$_2$D  phase  (see Fig~\ref{Tscan_SU3.fig} (a),(d)) the Ising transition  is again manifest in the sharp upturn in both the  dimer and the spin correlations.  For this value of the coupling both spin and dimer correlation grow below the Ising transition.

Overall the data is consistent  with a deconfined  phase at small values of $h$, a valence bond solid at intermediate values of $h$ and finally antiferromagnetic order at large values $h$.   The evolution of the order parameters is show in Fig.~\ref{Stag_mag_SU3.fig}(b).   From this we observe that
the SU(3) model not only captures a seemingly continuous transition between the $\mathbb{Z}_2$  Dirac deconfined and the VBS,  similar to the $N=2$ case, but also a transition from the spin-dimerized state (VBS)  to the SDW phase.     Since we observe no jump in the electric field and the order parameters also evolve continuously,  Eq.~\ref{Electric.eq}   leads to the conclusion that all transitions are continuous. In particular, the  VBS and  SDW phases is then expected to be an instance of a deconfined quantum critical point occurring  at $h \simeq 3/8$, similar to those found in quantum magnets \cite{Lou09}. The low energy theory for this transition has been argued to be three complex bosons coupled to a non-compact U(1) gauge field (non-compact CP${}_2$ theory) \cite{Senthil04_2}.  Within our limited system sizes we were not able to confirm  the literature value of the exponents \cite{Block13,Harada13}.  Note however that even for the largest   values of $L$ in Ref.~\cite{Harada13}    the $\nu$ exponent has still not converge. We  notice that  irrespective of the value of $\nu$, $\beta_{m}  < \beta_{VBS}$ where $\beta$ corresponds to the order parameter exponent. This inequality between $\beta_{m}$ and $ \beta_{VBS}$ is seemingly consistent with Fig.~\ref{Stag_mag_SU3.fig}(b) where VBS order parameter rises more sharply compared to the SDW as one moves away from the transition on the two sides.

\begin{figure*}
 \includegraphics[width=0.95\linewidth]{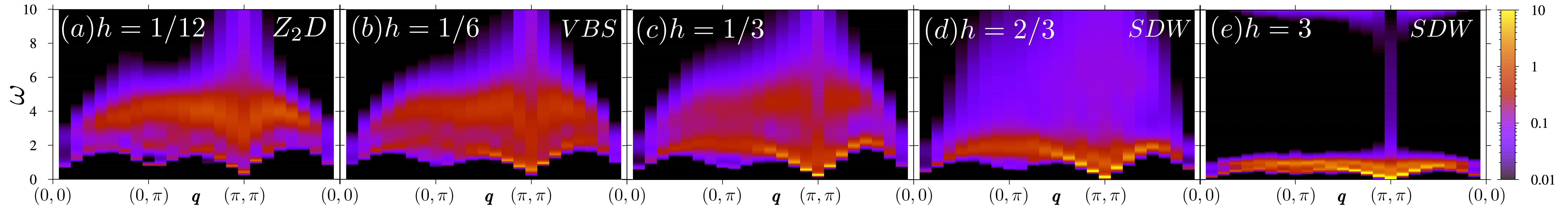} 
\caption{(color online)  Dynamical spin structure factor for the  SU(3) model.  The simulations were carried out on a $16 \times 16$ lattice  at $ \beta  = 30 $  for 
$h \geq 1/6$ and at  $ \beta = 60 $  for $h = 1/12$.  }
\label{Dyn_SU3.fig}
\end{figure*}

Fig.~\ref{Dyn_SU3.fig}   plots the evolution of the spin dynamical structure factor as a function of $h$.   As for the  SU(2) case, $\left[ \hat{Q}_{\pmb{i}},  \hat{S}^{\beta}_{ \; \; \alpha}(\pmb{j}) \right] =0$ such that intermediate states in the Lehmann representation of the zero temperature spectral function  do not violate the dynamically generated constraint.  
Since we are running  through two phase transition in rather close proximity,   we have to  disentangle features stemming   from the $\mathbb{Z}_2$ Dirac deconfined phase and from the  deconfined quantum critical point.  The characteristic of the Z$_2$D phase at $h = 1/12$ is the  spinon  continuum,   the high energy features  of which are very visible  across the  transition,  \textit{and} in the VBS  phase ($h=1/6$) but then fade away  when entering the SDW phase ($h \geq 3/8$)  and are completely absent  in the strong coupling limit at $h = 3$.  At this coupling strength,  one observes a pronounced Goldstone  mode in the vicinity of $\pmb{K} = (\pi,\pi)$  accounting for the  spin order.   Starting from this point and turning down the transverse field  $h$   results in a loss of weight of the spin wave mode. Upon close inspection, one equally sees that it shifts to higher energies.      

Another striking feature is the overall broadness of the spectral function  in the vicinity of the antiferromagnetic wave vector and in proximity of the VBS to SDW transition at $h \simeq 1/3$.  Qualitatively, this is consistent with a  deconfined quantum critical point which is characterized by a large value of  the anomalous dimension $\eta$ characterizing a complete breakdown of well defined quasiparticles  at criticality. Of course,  since the  spin susceptibility at criticality is expected to be of the form: 
\begin{equation}
	 \chi(\pmb{q},\omega) \simeq      \frac{1}{\left(   c^2|\pmb{q} -  \pmb{Q} |^2   - \omega^2  \right)^{1 - \eta/2}}
\end{equation}
with $c$ the spin velocity,   any non-zero $\eta$ implies absence of quasiparticles.   However at a conventional Wilson-Fisher fixed point in 2+1 dimensions, which describes a phase transition between an SDW and a featureless paramagnet in the Heisenberg universality class  $\eta \approx 0.036$ \cite{Wang06,Vicari02}, and correspondingly there are essentially no broad features visible in $S(\bm{q},\omega)$  \cite{Lohofer15}.  For the SU(3) J-Q model which shows a deconfined critical point,   $\eta \simeq  0.4$ \cite{Block13} . We note that the Z$_2$D to SDW transition observed in  the SU(2) model is equally characterized by a large value anomalous exponent   which within the $\epsilon$-expansion reads $\eta =  0.8$.  In the vicinity of the transition (see Fig.~\ref{Spin_Dynamics_SU2.fig} a) and b))   broad features at the  antiferromagnetic wave vector  are again apparent.

The data set at  $h = 3$  (see Fig.~\ref{Dyn_SU3.fig}e))   corresponds to the strong coupling limit where the spin wave velocity vanishes as $\frac{1}{2 N h }$.  Upon comparison between the data sets  of Fig.~\ref{Dyn_SU3.fig}d) and e)   one  notices a marked reduction of the spin wave velocity.  Another striking feature of the data, is the lack of  a sharp spin-wave mode as seen in the SU(2) case at our strongest  coupling (see Fig.~\ref{Spin_Dynamics_SU2.fig}e)). In fact  in the vicinity of the  $\pmb{q}=(\pi,0)$, one  detects  two distinct features.   We interpret this in terms of charge fluctuations around the mean-field  spin-density wave   state of Eq.~\ref{NeelSU3.eq}.  Setting   $\alpha = 1$  in this equation  and applying Hamiltonian  $\hat{H}_{\infty} $ will generate  spin-flip excitations on bonds with $\alpha = 2,3$.  This set of excitations lead to the spin-density wave mode.   One will however equally   generate charge fluctuations  where  one site  of the bond is left empty and  the other accommodates three particles.  We interpret the additional features  observed in the data in terms of these charge fluctuations. At $h = \infty$    spin and charge excitations are degenerate and the spin wave velocity as well as the  energy cost of the charge excitation are tied to the same scale $\frac{1}{2 N h }$ such that we cannot dissociate both features.   Furthermore one easily sees that such charge fluctuations do not occur in the SU(2) case, where we observe a sharp  spin-wave mode (Fig.\ref{Spin_Dynamics_SU2.fig}e).

\section{Summary and Future Directions}

Perhaps the most attractive feature of our model (Eq.\ref{Model.Eq}) is that despite its simplicity and the absence of fermion sign problem, the phase diagram is rather rich (Fig.\ref{Phase.fig}) and features several highly entangled phases and quantum phase transitions. In particular, in the $\mathbb{Z}_2$ deconfined phase, topological degrees of freedom, namely, the dynamically generated $\mathbb{Z}_2$ gauge field is coupled to gapless fermions. From an entanglement perspective,  the $\mathbb{Z}_2$ Dirac phase will  result in sharp universal footprints in their entanglement structure and this provides an attractive opportunity to apply some of the recently developed techniques to calculate entanglement entropies and entanglement spectrum \cite{Grover13,Assaad13a,Broecker14,Assaad15,Drut15}. Specifically, the universal part of the entanglement in the $\mathbb{Z}_2$ Dirac phase is expected to follow the relation $\gamma_{Z_2 D} = \gamma_{D} + \log(2)$ where $\gamma_{D}$ is the universal value for the same shaped region in a regular Dirac phase \cite{Yao10}. It would be similarly interesting to study the entanglement structure at the transition from Z$_2$D to symmetry broken phases for $N=2$ and $N=3$, as well as at the deconfined critical point between SDW and VBS.

The transitions from Z$_2$D phase to the SDW/SC for $N=2$ and to VBS for $N=3$ are rather exotic since they appear to be continuous while involving both symmetry breaking, and confinement of the gauge field \footnote{We thank Ashvin Vishwanath for discussion on this point}. We don't have an understanding of these transitions and it's a worthwhile direction to pursue.

Although our model is not microscopically motivated by any specific material(s), the resulting phenomenology does share several features with spin-liquid candidate materials \cite{Kagawa05, Han12, Shimizu03, Itou08} which also exhibit broad, non-quasiparticle features in their excitation spectrum. It is also worth noting that there have been recent proposals for realizing dynamical gauge-matter theories  in cold atomic lattices where instead of imposing the gauge constraint as a penalty term, the Hamiltonians such as ours  (Eq.\ref{Model.Eq}) appear naturally due to angular momentum conservation \cite{Zohar13, Zohar16}.

Our model  features an interesting interplay of several symmetries -- an  extensive number of conserved operators  $\hat{Q}_{\pmb{i}}$ that anti-commute with the particle-hole symmetry, as well as continuous O(2N) symmetry that allows one to perform sign free Monte Carlo for all N. The particle-hole symmetry is the reason that our model hosts a finite temperature transition where $\hat{Q}_{\pmb{i}}$ spontaneously orders leading to an effective `Gauss's law' for a $\mathbb{Z}_2$ gauge-matter theory, unlike a conventional gauge theory \cite{Kogut75, Senthil00} where the Gauss's law  is imposed explicitly on the Hilbert space. Contrast this also with 2D Toric code \cite{Kitaev03} where the Gauss's law is imposed energetically by explicitly adding a term that is proportional to $\sum_{\bm{i}} \hat{Q'}_{\pmb{i}} = \sum_{\bm{i}}  \hat{X}_{\pmb{i},\pmb{i} + \pmb{a}_x} \hat{X}_{\pmb{i},\pmb{i} - \pmb{a}_x} \hat{X}_{\pmb{i},\pmb{i} + \pmb{a}_y} \hat{X}_{\pmb{i},\pmb{i} - \pmb{a}_y}$. Since this term breaks the  $\hat{Q'}_{\pmb{i}} \rightarrow -  \hat{Q'}_{\pmb{i}}$ symmetry explicitly, there is no finite temperature transition in the 2D Toric code. The fields $\hat{Q}_{\pmb{i}}$ in our model are of course classical in the sense that they are conserved and don't have any dynamics. It would be desirable to extend the idea of studying deconfined phases of gauge theories without imposing the gauge constraint to other gauge groups as well, particularly to the case of $U(1)$ gauge theory relevant to several spin-liquid candidates. Imposing gauge constraint numerically is often challenging, and our results indicate that at least in certain gauge theories, is not necessarily required. 

On the above note, it would also be interesting to explore Toric code \cite{Kitaev03} like Hamiltonians with inbuilt $\hat{Q'}_{\pmb{i}} \rightarrow -  \hat{Q'}_{\pmb{i}}$ symmetry, and study the implications of the melting of the Gauss's law at finite temperature. This could be especially interesting in 4D Toric code where there exist finite temperature quantum memory \cite{Dennis02}. Is the quantum memory lost as soon as  the thermal expectation value $\langle \hat{Q'}_{\pmb{i}}\rangle$ vanishes?

In our models, the finite temperature transition corresponding to the ordering of $\hat{Q}_{\pmb{i}}$ coincided with spontaneous breaking of particle-hole symmetry  for all values of $N$. This is because $\langle \hat{X}_{\pmb{i},\pmb{i} + \pmb{a}_x} \hat{X}_{\pmb{i},\pmb{i} - \pmb{a}_x} \hat{X}_{\pmb{i},\pmb{i} + \pmb{a}_y} \hat{X}_{\pmb{i},\pmb{i} - \pmb{a}_y} \rangle \neq 0$. It is interesting to contemplate the possiblity of a phase where this is not the case such that $\hat{Q}_{\pmb{i}}$ order while the spatial correlation functions of $\hat{p}_{\pmb{i}} $  are short-ranged. Consider for example the model:

\begin{equation}
	\hat{H}=\sum_{\left< \pmb{i},\pmb{j} \right> }   \left[ \hat{Z}_{\langle \pmb{i},\pmb{j} \rangle}  \left( \hat{c}^{\dagger}_{\pmb{i}} \hat{c}_{\pmb{j}}  + \text{H.c.} \right) +  u \, \hat{Q}'_{\pmb{i}} \hat{Q}'_{\pmb{j}} \right] + v \sum_{\pmb{i}} \hat{Q}_{\pmb{i}} \label{eq:altH}
 \end{equation}
where $ \hat{Q}'_{\pmb{i}} = \hat{X}_{\pmb{i},\pmb{i} + \pmb{a}_x} \hat{X}_{\pmb{i},\pmb{i} - \pmb{a}_x} \hat{X}_{\pmb{i},\pmb{i} + \pmb{a}_y} \hat{X}_{\pmb{i},\pmb{i} - \pmb{a}_y} $. One can implement a  $\hat{Q}'_{\pmb{i}} \rightarrow -  \hat{Q}'_{\pmb{i}}$ symmetry in the above Hamiltonian by demanding that $\hat{H}$ be invariant under the transformation $\hat{U}^{-1} \hat{H} \hat{U}$ where $\hat{U} = \prod'_{\langle \pmb{i},\pmb{j} \rangle}  \hat{Z}_{\langle \pmb{i},\pmb{j} \rangle} \prod_{{\pmb{i}}} \hat{P}_{\pmb{i}}$ where the prime on the first product denotes that it is taken over all vertical (horizontol) links either only for even rows (columns) or only for odd rows (columns). This transformation takes $\hat{Q}'_{\pmb{i}} \rightarrow -  \hat{Q}'_{\pmb{i}}$, $\hat{Q}_{\pmb{i}} \rightarrow \hat{Q}_{\pmb{i}}$ and $\hat{X}_{\pmb{i},\pmb{j}} \rightarrow - \hat{X}_{\pmb{i},\pmb{j}} $. Therefore, even though $\langle \hat{Q}_{\pmb{i}} \rangle \neq 0$ in this model, the order parameter for the charge-ordering, namely,  $\langle \hat{p}_{\pmb{i}} \rangle$ will be non-zero only if the $\hat{U}$ symmetry is spontaneously broken. A phase where $\hat{Q}_{\pmb{i}}$ order while $\hat{p}_{\pmb{i}}$ remain disordered would necessarily involve interesting entanglement between the fermions and the Ising degrees of freedom because one requires that $\langle \hat{X}_{\pmb{i},\pmb{i} + \pmb{a}_x} \hat{X}_{\pmb{i},\pmb{i} - \pmb{a}_x} \hat{X}_{\pmb{i},\pmb{i} + \pmb{a}_y} \hat{X}_{\pmb{i},\pmb{i} - \pmb{a}_y} \hat{p}_{\pmb{i}}\rangle   \neq 0$ while $\langle \hat{p}_{\pmb{i}} \rangle = 0$.

The $N=3$ model has the richest phase diagram among the models we study. In particular, apart from the continuous  transition between Z$_2$D and VBS, the phase diagram contains two distinct symmetry broken phases, the VBS and the SDW phase. Remarkably, we find that these two phases are separated by a second order transition (Fig.~\ref{Stag_mag_SU3.fig}b), which is thus expected to belong to  a non-compact CP${}_2$  universality class \cite{Senthil04_2}. At this deconfined quantum critical point, we also study the imaginary-time dynamics via the calculation of structure factor $S(\bf{q}, \omega)$(Fig.~\ref{Dyn_SU3.fig}, and found a rather broad continuum of excitations, consistent with the theoretical expectation that  at this transition, the effective description is in terms of fractionalized `spinons'. To our knowledge, the dynamics at the deconfined quantum criticality has not been studied numerically in the past, and given rather distinct features seen in our work, it would be worthwhile to pursue this direction in quantum spin-models where much larger sizes are accessible via bosonic Quantum Monte Carlo such as Stochastic Series Expansion (SSE) \cite{Sandvik07}.

Some cautionary statements are in order for the $N=3$ case. We found it challenging to perform scaling collapse at the both quantum phase transitions, despite no indication of a first-order transition in the behavior of the order parameter. One reason for this could be that the two zero temperature transitions are rather close by in the phase diagram. This is visible in the behavior of the structure factor as $h$ is tuned (Fig.~\ref{Dyn_SU3.fig}) where a diffuse continuum of excitations exists all the way from $h = 0$ ($\mathbb{Z}_2$ Dirac phase) to $h \sim 1/3$ (VBS to SDW transition). It's also worth mentioning that the system sizes we are working ($L \leq 16$), are an order of magnitude smaller than the ones for spin-models which exhibit deconfined quantum criticality.

We notice that when $N$ is even, our model does not suffer from sign problem even in the presence of the a non-zero chemical potential on arbitrary lattices. Thus it would be interesting to extend our study to a system with finite density of fermions.

We would like to conclude with a conjecture only tangentially related to the rest of the paper. Consider the Hamiltonian in Eq.~\ref{h_inf.eq} for $N=2$ on an arbitrary graph, $\hat{H}_{\infty} = \hat{H}_{\bm{\eta}} + \hat{H}_{\bm{\sigma}}$ where  $\hat{H}_{\bm{\eta}} = \sum_{ \langle \pmb{i},\pmb{j} \rangle } J_{\pmb{i},\pmb{j}} \left( \hat{\eta}_{\pmb{i}}^{z} \hat{\eta}_{\pmb{j}}^{z} - \frac{1}{2}\left(\hat{\eta}_{\pmb{i}}^{+} \hat{\eta}_{\pmb{j}}^{-} + \hat{\eta}_{\pmb{i}}^{-} \hat{\eta}_{\pmb{j}}^{+}\right) \right)$ and $\hat{H}_{\bm{s}} =
\sum_{ \langle \pmb{i},\pmb{j} \rangle } J_{\pmb{i},\pmb{j}} \left( \hat{s}_{\pmb{i}}^{z} \hat{s}_{\pmb{j}}^{z} + \frac{1}{2}\left( \hat{s}_{\pmb{i}}^{+} \hat{s}_{\pmb{j}}^{-} + \hat{s}_{\pmb{i}}^{-} \hat{s}_{\pmb{j}}^{+}\right)\right)  $. Assuming $J_{\pmb{i},\pmb{j}} \geq 0$, the Hamiltonian $\hat{H}_{\infty}$ is  sign problem free on an arbitrary graph because one can use Hubbard-Stratanovich decomposition to write the partition function as a sum of non-negative fermion determinants. On the other hand, considering its decomposition into spin-Hamiltonians $\hat{H}_{\bm{\eta}}$ and  $\hat{H}_{\bm{s}}$, only $\hat{H}_{\bm{\eta}}$  is sign problem free on an arbitrary graph due to all non-positive off-diagonal elements in a local basis. Since $\left[\bm{\eta},\bm{S}\right]=0$, the simulatability of $\hat{H}_{\infty}$ suggests that its ground state is likely identical to that of $\hat{H}_{\bm{\eta}}$.  That is, in the ground state one expects no singly occupied sites -- half of the sites are unoccupied while the other half are doubly occupied, so that only $\hat{H}_{\bm{\eta}}$ acts non-trivially. This leads to the following conjecture: \textit{Given a set of $J_{ij} \geq 0$, the ground state energy of} $\hat{H}_{-} = \sum_{ \langle \pmb{i},\pmb{j} \rangle } J_{\pmb{i},\pmb{j}} \left( \sigma_{\pmb{i}}^{z} \sigma_{\pmb{j}}^{z} - \frac{1}{2}\left(\sigma_{\pmb{i}}^{+} \sigma_{\pmb{j}}^{-} + \sigma_{\pmb{i}}^{-} \sigma_{\pmb{j}}^{+}\right)\right)  $ \textit{ is less than or equal to the ground state energy of} $ \hat{H}_{+} = \sum_{ \langle \pmb{i},\pmb{j} \rangle } J_{\pmb{i},\pmb{j}} \left( \sigma_{\pmb{i}}^{z} \sigma_{\pmb{j}}^{z} + \frac{1}{2}\left(\sigma_{\pmb{i}}^{+} \sigma_{\pmb{j}}^{-} + \sigma_{\pmb{i}}^{-} \sigma_{\pmb{j}}^{+}\right)\right)  $. Clearly, on a bipartite graph this conjectured inequality is always saturated. To test this conjecture, we performed exact diagonalization study  for systems upto 10 qubits when $J_{ij}$ are chosen randomly from a uniform distribution on $[0,1]$ for a sample size of 1000. We didn't find any counterexample.

Prior to submission of this paper, we were aware of related work by Snir Gazit, Mohit Randeria and Ashvin Vishwanath \footnote{Snir Gazit, Mohit Randeria and Ashvin Vishwanath, arXiv (to appear)}. We thank them for sharing their draft.

\noindent
\textbf{\underline{Acknowledgements:}}   
We would  like to thank Yin-Chen He, I. Herbut, A. L\"auchli,  F. Parisen Toldin,  S. Trebst, Mike Zaletel  and especially Tim Hsieh for discussions. We especially thank Ashvin Vishwanath for pointing out  discrepancies in our understanding of the transition between the  Z$_2$D  and SDW phase   in an earlier draft of the paper.  

TG acknowledges startup funds from UCSD.  FFA   is supported by the German Research Foundation (DFG), under  the  DFG-SFB 1170  ToCoTronics  (Project C01). We thank the J\"ulich Supercomputing Centre for generous allocation of CPU time.  This research was supported in part by the National Science Foundation under Grant No. NSF PHY11-25915.

\end{document}